\documentclass{jaa}

\usepackage{graphicx}
\usepackage{lscape}

\begin{document}\sloppy

\title{Planetary Nebulae with UVIT: A Progress Report.}


\author{N. Kameswara Rao\textsuperscript{1,*}, Sutaria F. \textsuperscript{1}, Murthy J.\textsuperscript{1},  Ray A.\textsuperscript{2,3} \and Pandey G. \textsuperscript{1} }
\affilOne{\textsuperscript{1}Indian Institute of Astrophysics, Bangalore 560034, India. \\}
\affilTwo{\textsuperscript{2}Tata Institute of Fundamental Research, Colaba, Mumbai 400005,India. \\}
\affilThree{\textsuperscript{3} Homi Bhabha Centre for Science Education (TIFR) Mumbai 400088, India.}


\twocolumn[{

\maketitle

\corres{nkrao@iiap.res.in}

\msinfo{7 Nov. 2020}{--}

\begin{abstract}
  The spectral region between 1250 $\AA$ -3000 $\AA$  contains important spectral
           lines to understand the morphological structures and evolution of planetary
           nebulae. This is the region sampled by UVIT through various filter bands
           both in the continuum and in emission lines (e.g.. C\,{\sc iv}, He\,{\sc i},
           Mg\,{\sc ii} etc.).
           We have mapped several planetary nebulae with different characteristics,
           ranging in morphology from bipolar to wide and diffuse, and in various states of
           ionization, comparing the UV with the x-ray morphologies wherever the x-ray images
           were also available. The major unanticipated discovery with UVIT has been the detection of 
           previously undetected, cold, fluorescent, H$_2$ gas surrounding some planetary nebulae.
           This may be a possible solution to the missing mass problem. Here we present a review of 
           our studies so far done (both published and on going) with UVIT.
\end{abstract}

\keywords{Star: AGB and post --- AGB stars, winds, outflows --- planetary nebulae: ISM: Planetary nebulae: general --- planetary nebulae: individual: NGC 6302.}
}]


\doinum{12.3456/s78910-011-012-3}
\artcitid{\#\#\#\#}
\volnum{000}
\year{0000}
\pgrange{1--}
\setcounter{page}{1}
\lp{1}

\section{Introduction}
              Planetary nebulae (PNs) are splendid remnants of extraordinary deaths of
   ordinary stars in the mass range of 1 -- 8 M$_{\rm \odot}$. They disburse the 
  nucleosynthetically processed stellar material like Carbon and s-process elements into the interstellar medium, thus enriching
the matter which forms the next generation of stars. The extensive, slow, stellar wind, moving at speeds of 10 to 15 km s$^{-1}$,  
with a mass-loss  rate of  $\sim 10^{-7}$ M$_{\rm \odot}$ yr$^{-1}$, that starts on the thermally pulsing asymptotic giant branch (AGB) 
-- double shell  (He \& H)  burning sources -- transforms  into a heavy super-wind with mass-loss rates of  $\sim 10^{-4}$ M$_{\rm \odot}$ yr$^{-1}$
  (Delfosse et al. 1997)  as the star evolves to the tip of AGB  in the H-R diagram. 
   In a relatively short time most of the mass is lost through a super-wind till the  
  envelope mass falls below  $10^{-3}$ - $10^{-4}$ M$_{\rm \odot}$,  when  a structural 
  change occurs to the star as a degenerate CO oxygen core (which ultimately becomes a white dwarf) develops. 
  The photospheric radius shrinks and the effective temperature  T$_{\rm eff}$ starts to increase  keeping
  the luminosity almost constant. Consequently, the mass-loss rate stellar wind decreases to about
  $10^{-8}$ M$_{\rm \odot}$ yr$^{-1}$ and the wind speed picks up to 200 to 2000 km s$^{-1}$.
   This fast stellar wind plows into the material that was earlier lost through super-wind 
 generating a shock at the interface, while the stellar radiation heats and ionizes the ejecta.
   The circumstellar material starts to glow as the planetary nebula, and ``illuminates the pages of the book that tells 
  the star's story'' (Bianchi 2012). In the interacting stellar wind model (Kwok 1978), it the interaction of the the high speed stellar
  wind and the slowly expanding super-wind material that shapes the planetary nebulae. Presence of a companion and or 
  magnetic fields may further alter the morphology of the PN. In general, PNe show very many shapes
  ranging from spherical to bipolar to  multipolar, with some even having chaotic geometries. Morphological studies
   of these objects reveal their past history of mass ejections, their time scales,
   kinematics,  properties of the ionizing source, wind interactions as well as
  interactions with interstellar medium etc.. The UV region is important for the study
   of both the central stars (CSPNs) as well as the nebula, because the most important 
   lines of the most abundant elements and their ionization states like C\,{\sc ii} 1335 \AA, C\,{\sc iv} 1550 \AA, 
   He\,{\sc ii} 1640 \AA,N\,{\sc iii}] 1760 \AA C\,{\sc ii}] 2326 \AA etc., fall in this region. These lines are important 
   for modeling the ionization structure, shocked regions, chemical composition etc., and for the estimation of the 
   T$_{\rm eff}$ of the hot CSPNs. Moreover, the interstellar extinction through 2179 \AA bump can be studied only in the UV band.
   The Ultraviolet Imaging Telescope(s) (UVIT) on ASTROSAT (ref.), with broad and narrow band filters which cover important spectral 
   lines and continuum with an angular resolution of about 1".5, over a 28' field of view, are well suited for the study
   of PNs.

          Details of UVIT are provided in Kumar et al. (2012) and its in-orbit
   performance is described in Tandon et al. (2017a) and in Tandon et al. (2020). UVIT is one of the five payloads on the
   multiwavelength Indian astronomical satellite ASTROSAT that was launched on 2015 September 28.
   It consists of two 38 cm aperture telescopes, one of which is  optimized for
   FUV, while  the other has a dichroic beam splitter that reflects NUV and transmits
   the optical. Each UV channel can be studied in five broad and narrow band filters, as well
   as by low resolution transmission gratings. The Visual channel (VIS channel), which operates only in the 
   integration mode, is used for tracking. Our project uses UVIT imaging of x-ray bright and x-ray faint Planetary Nebulae of different morphologies
   in various UV emission lines, particularly C\,{\sc iv} 1550 \AA,
   C\,{\sc ii}] 2326 \AA, [O\,{\sc ii}] 2470 \AA, Si\,{\sc iv} 1400 \AA, Mg\,{\sc ii}] 2800 \AA,
   He\,{\sc ii} 1640 \AA\  etc.,   using various filters of the UVIT-FUV and -NUV channels. We aim
    to study the UV morphologies, shocked regions and correspondence of
   UV and X-ray emissions in PNs, and to that end, several PNs of varied morphological types
   in both near (NUV) and far (FUV) UV ranges have been observed. Unfortunately, the NUV channel became dysfunctional
   after 2017. In this paper, we discuss our observations conducted so far of the selected
    PNs  (Table 1), as well as  some of the results and surprises that emerged.
   Detailed studies of individual objects would be presented else where but some salient 
   observational features particularly brought out by UV studies are dealt in the current
   presentation. Detailed discussion of NGC 40 and NGC 6302 have been presented in
   Kameswara Rao (2018a,b).

                     Table 1 shows a broad  morphological classification of the nebulae we observed
  with UVIT so far which range from compact bipolar nebulae (B) to large elliptical (E) and
  round (R) nebulae. Some are irregular. Figure 1 illustrates typical nebular  emission lines
  that are enclosed by  UVIT filters that were used for our PN studies.

\begin{figure*}
\vspace{0.0cm}
\includegraphics[width=18cm,height=15cm]{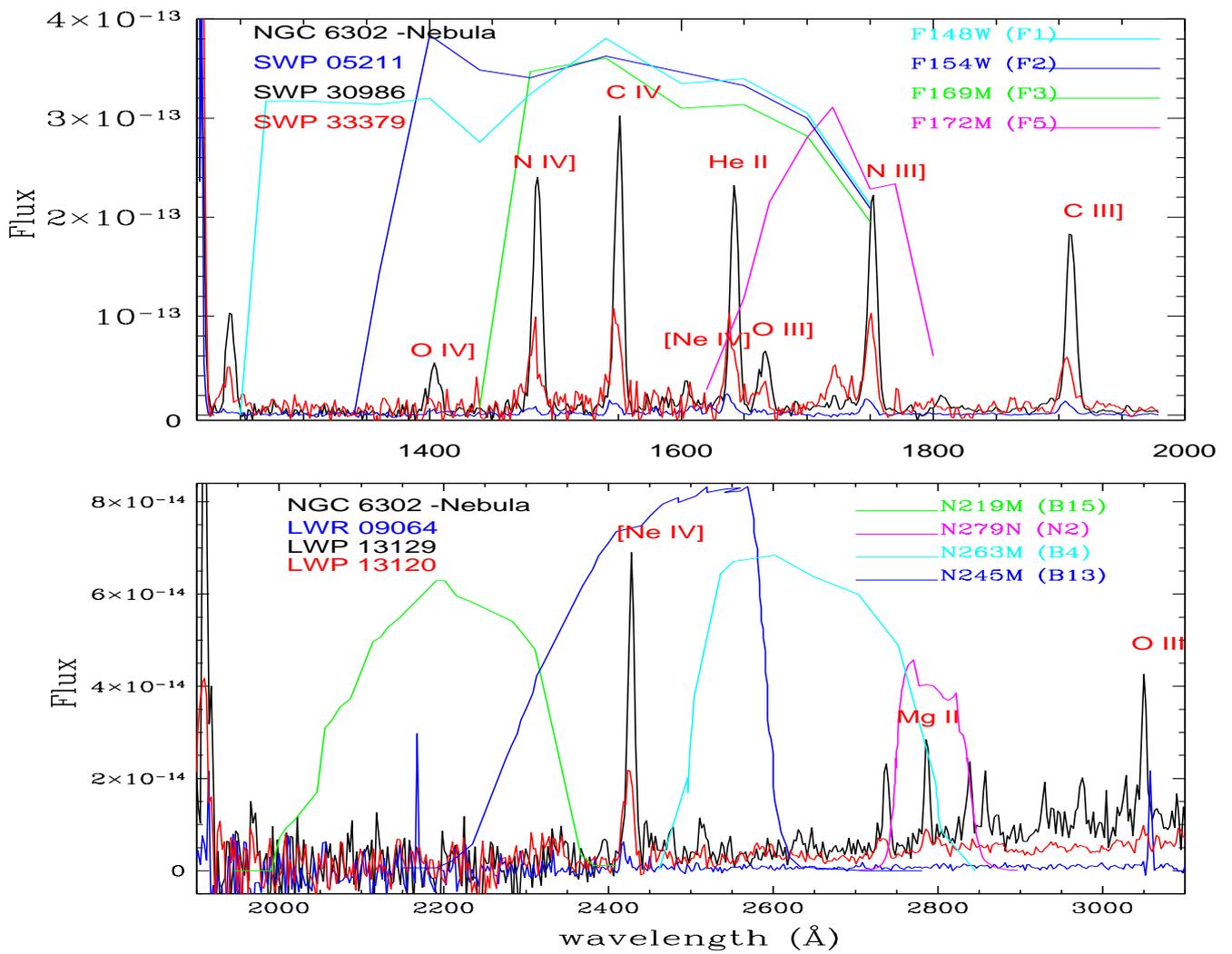}
\caption{ IUE low resolution nebular spectra of NGC6302 are shown to
  illustrate the wavelength range of UVIT filters. FUV is
  plotted on top and   NUV at the bottom.
  Relative effective areas of various UVIT filters used for PN studies
  and typical nebular emission lines they include are shown. }

\end{figure*}
 \section{Results and Discussion}

\subsection{Bipolar Nebulae -FUV Halos and Arcs}

             Most of the nebulae we have observed so far belong to the group of PNe with bipolar
morphology. Our sample includes  NGC 40 (the "Bow tie nebula"), NGC 650, NGC 2440, OH231.8+4.2 (the "Calabash nebula"), NGC 2818, Mz 3,
 NGC 6302 and NGC 7027. One of the reasons we observed them is to look for systematic featufres (aspects) in the UV that would characterise the group -
  e.g. the cold circumnebular H$_2$ gas. Detail UVIT imaging studies of NGC 40 and NGC 6302 are to be found in (Kameswara Rao et al. 2018a,b), while  
  and for NGC 2818 in (Kameswara Rao et al.,  A\& A, submitted). Although observations of NGC 650, OH231.8+4.2, Mz3, (and IC 4997) have been done
 the data is not yet available from ISSDC.

The compact low excitation planetary nebula, NGC 40, was the first object we studied with a view to look for correspondence of high excitation UV line 
regions with Chandra X-ray images. It has been imaged in the far-ultraviolet filters F169M (UVIT/FUV-F3  with $\lambda_{\rm eff} = 1608$ {~\AA} ) and F172M (UVIT/FUV-F5 with $\lambda_{\rm eff}$ of {1717~\AA}), as well as in the near-ultraviolet (UVIT/NUV) filters N245M (UVIT/NUV-B13) and N279N (UVIT/NUV-N2 with $\lambda_{\rm eff}$ of {2792~\AA}).
 The filters selected  would allow imaging in  C\,{\sc iv} 1550\AA\ (F169M) and  C\,{\sc ii}] 2326\AA\ (N245M) emission lines, as well as in the continuumn (F172M) and (N263). Morphological studies in optical and infrared (IR) show that  NGC 40 has ionized high density central core surrounded by faint filamentary halo with circumnebular rings that are seen only in H$\alpha$ but not in [O\,{\sc iii}].
  UVIT studies show that C\,{\sc ii}] 2326\AA\ emission is confined mostly to the core and
  shows similar morphology as low excitation lines in optical. However,
  strong C\,{\sc iv} 1550\AA\ emission is present in the core and shows similar morphology and extent as that of X-ray (0.3-8 keV) emission 
  observed by {\it Chandra}, suggesting
   interaction of the high-speed wind from WC8 central star (CS) with the nebula.
   An unexpected UVIT discovery is the presence of faint large emission halo in FUV F169M
  surrounding the central core (Figure 2-top). This FUV halo is absent in the other filters. This
   emission halo is unlikely to be due to  C\,{\sc iv} 1550\AA\ emission, or due to dust scattering. Instead, 
  it most likely is due to UV fluorescence emission from Lyman bands of H$_2$ molecules since a few vib-rotational lines have already been
  detected in the IR from Spitzer spectra.
  The FUV halo in NGC 40 highlights the extensive existence of cold H$_2$ molecules in the regions even
  beyond the optical and IR halos. Thus UV studies are important to estimate the amount of H2,
  which is probably the most dominant molecule and significant for mass-loss studies.
  Central star and the nebular core occur in the north-west edge of the FUV halo in the direction
  of the star's proper motion vector suggesting a possible interaction with the surrounding
   interstellar medium (ISM).

\begin{figure*}
\vspace{0.0cm}
\includegraphics[width=17cm,height=11cm]{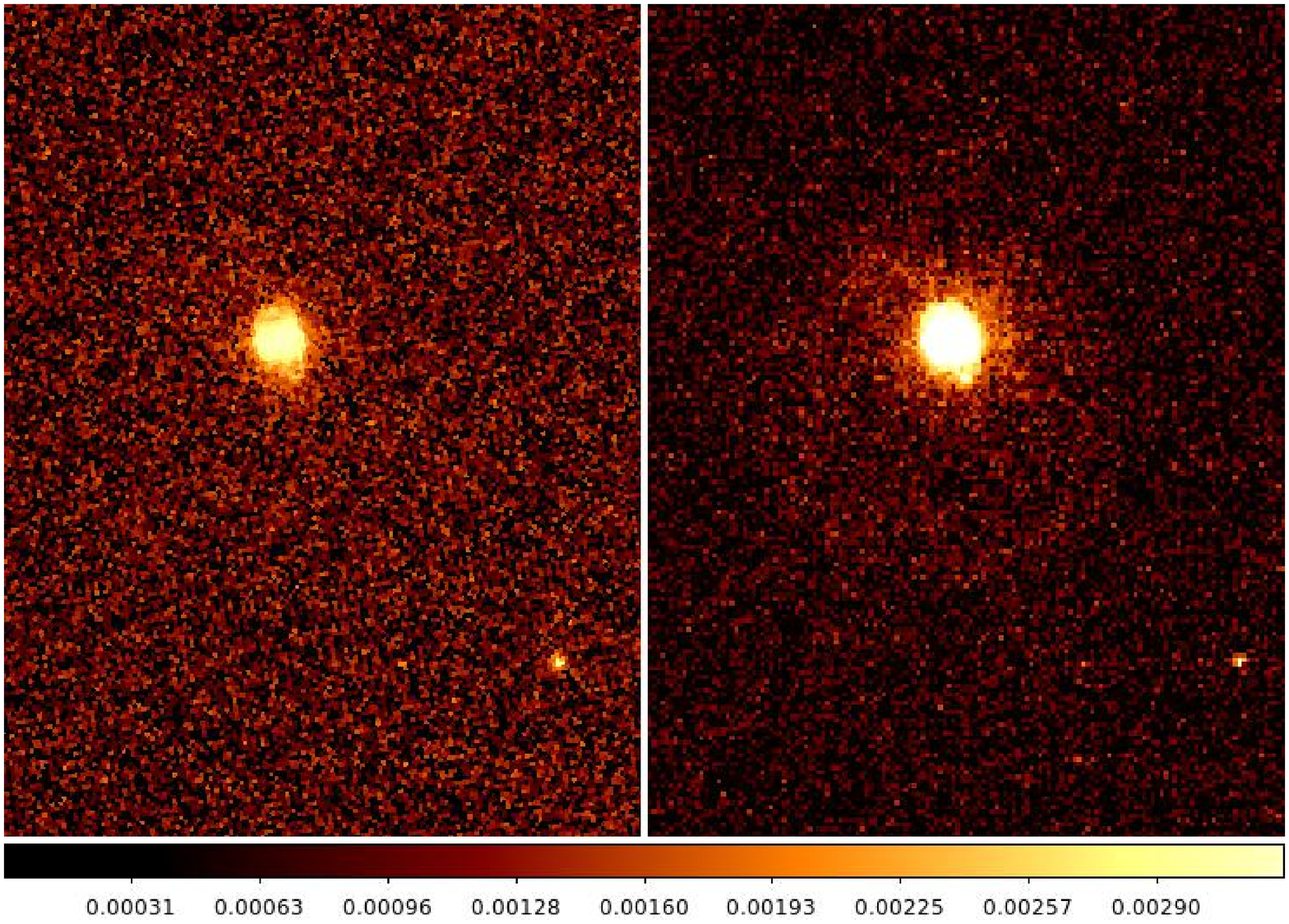}
\includegraphics[width=17cm,height=11cm]{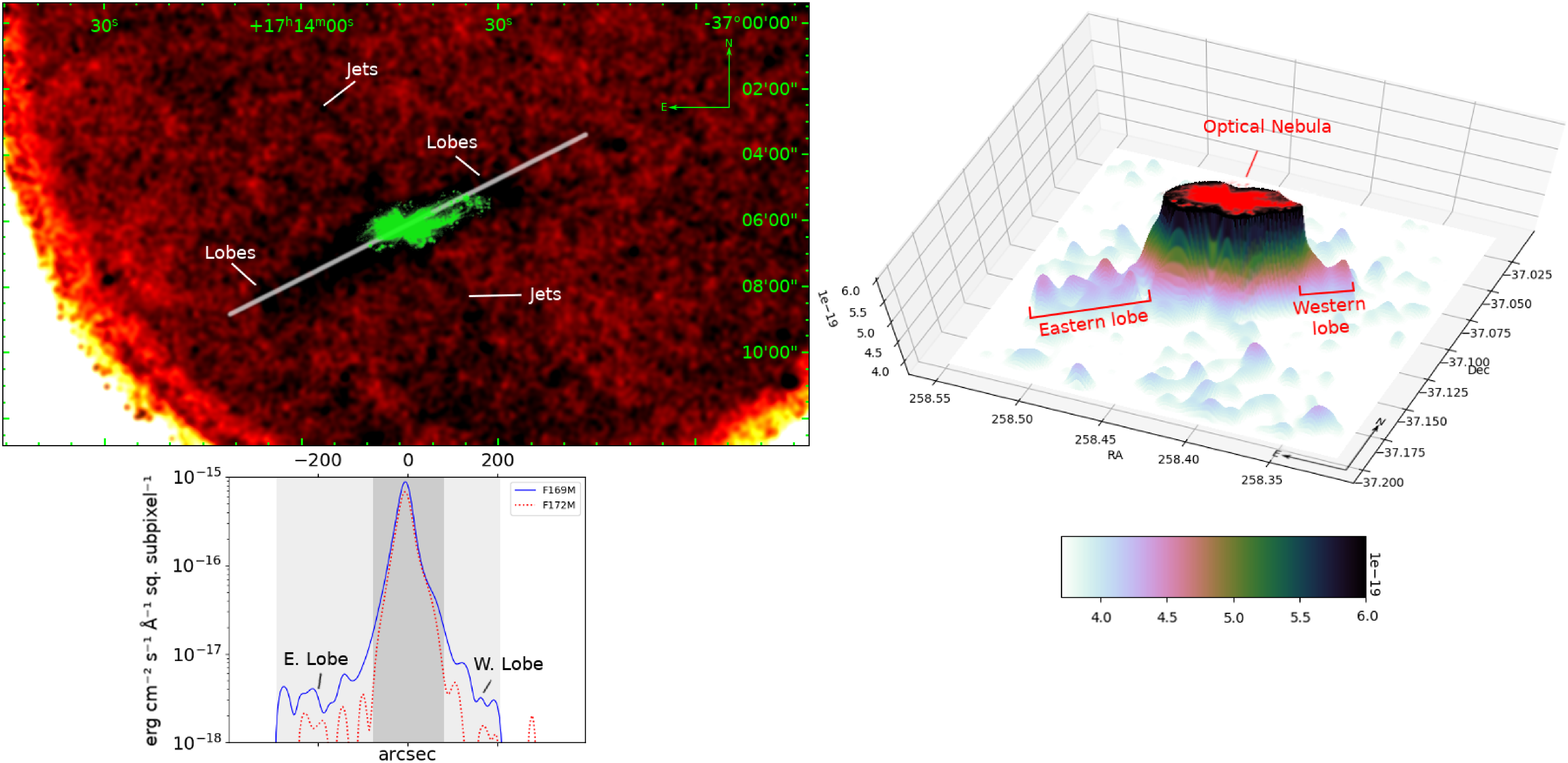}
\caption{(Top:) FUV images of NGC 40 in F172M (left) and in F169M (right). The faint FUV halo
 in F169M,  extending beyond the bright central region, is absent in the F172M image. 
  (Kameswara Rao et. al 2018a). 
 (Bottom:) The extensive FUV lobes and jets in NGC 6302 are shown in the F169M image which
  extends much beyond the optical image. The F172M image (not shown) does not show these lobes
  and jets.  }
\end{figure*}

         Presence of much bigger and more extensive FUV halo was discovered around the famous
  high excitation PN, NGC 6302 the butterfly nebula (Kameswara Rao et al. 2018b). It  has been
  imaged in  F169M  and F172M, as well as in N279N  and in N219M (UVIT/NUV-B15 with  $\lambda_{\rm eff}$ of {2196~\AA}).

\begin{figure*}
\vspace{0.0cm}
\includegraphics[width=10cm,height=8cm]{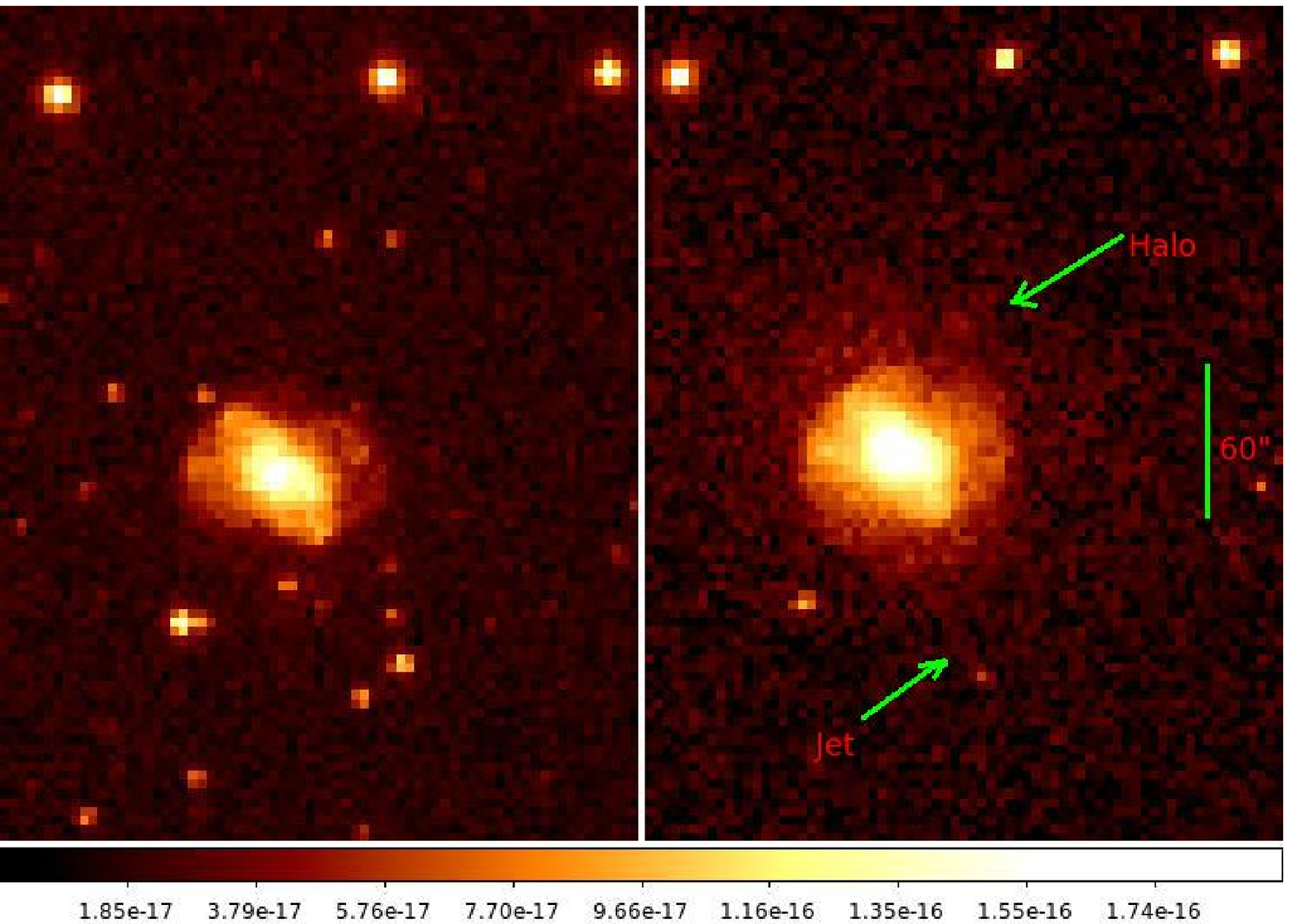}
\caption{(Top) :NUV N219M image (left) is compared with UVIT image
 in F169M (F3) of NGC 2440.  }
\end{figure*}

\begin{figure*}
\vspace{0.3cm}
\includegraphics[width=8.0cm,height=7cm]{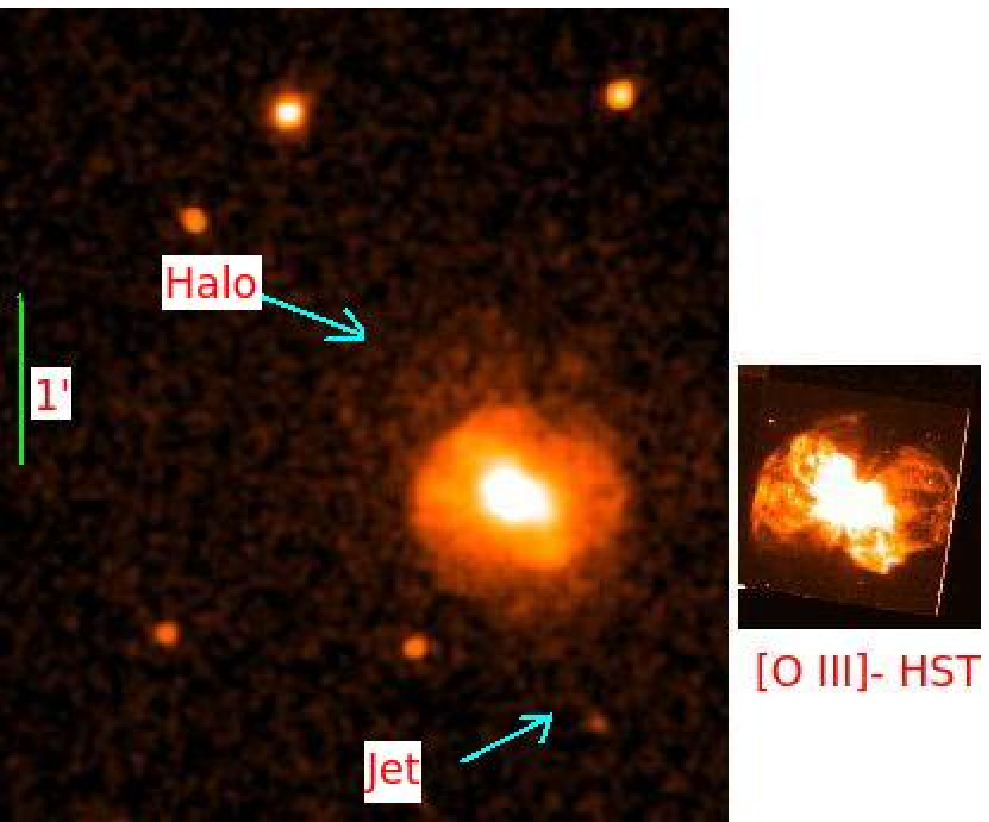}
\vspace{0.3cm}
\includegraphics[width=8.0cm,height=7cm]{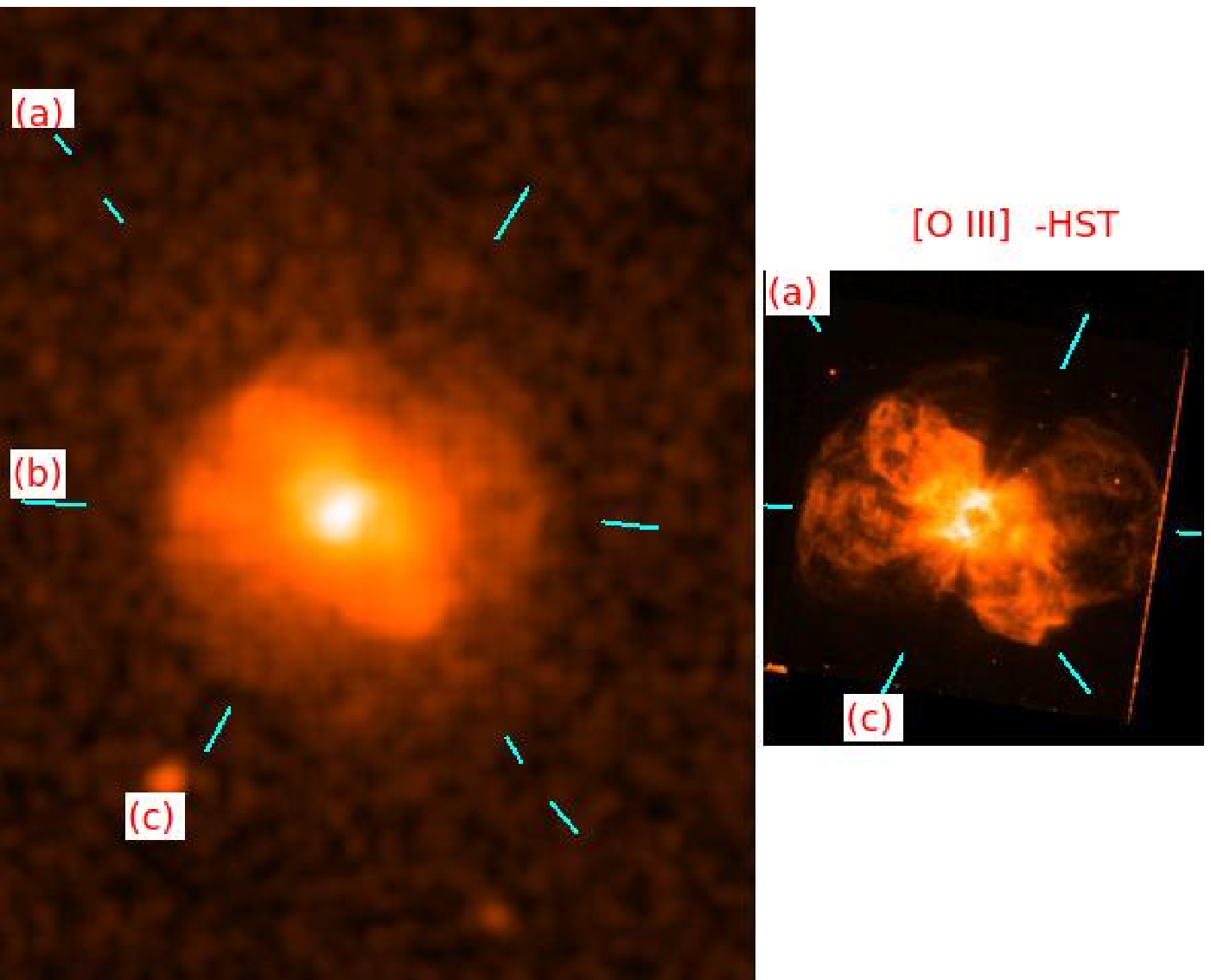}
\vspace{0.3cm}
\includegraphics[width=8.5cm,height=6.5cm]{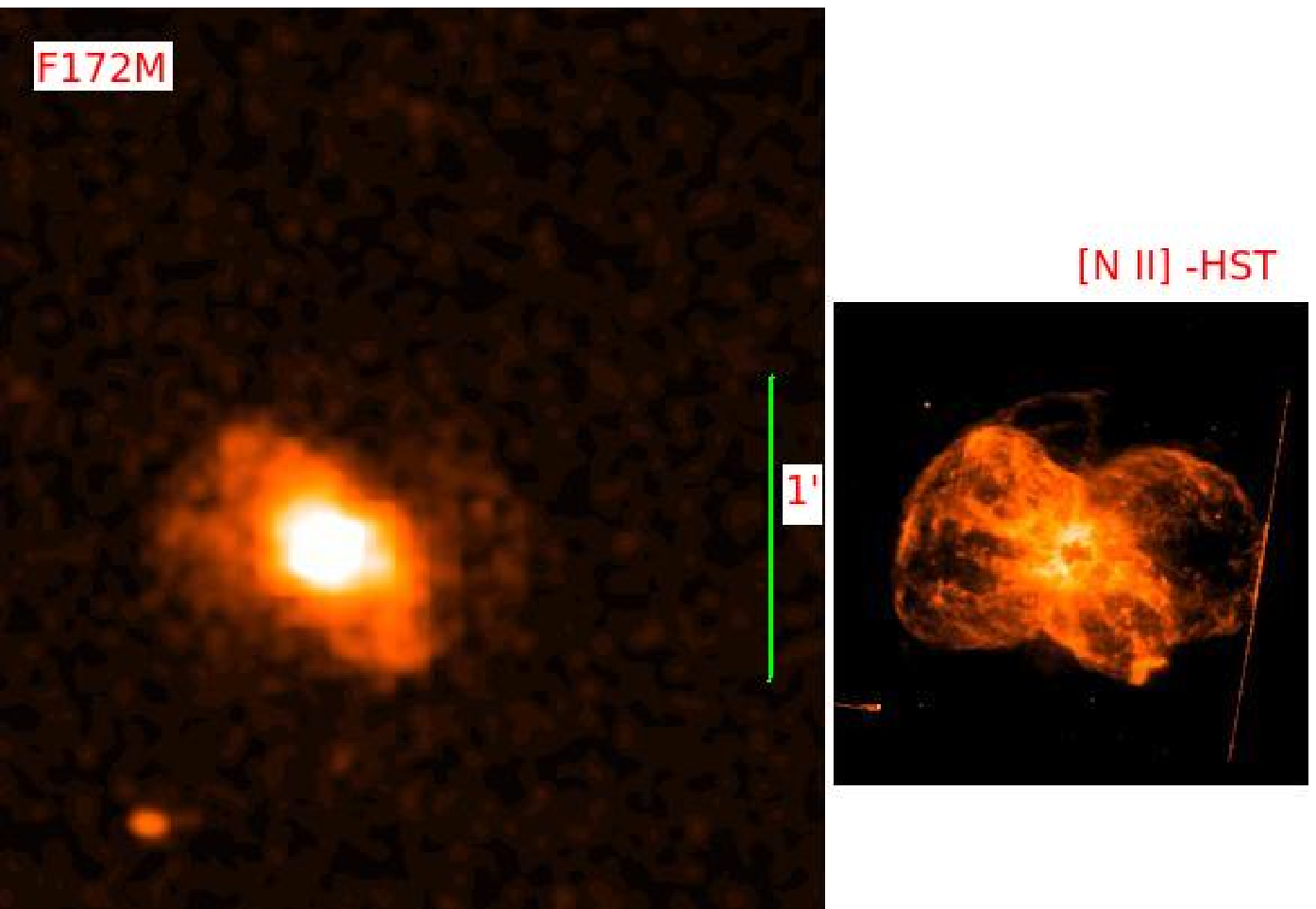}
\vspace{0.3cm}
\includegraphics[width=8.5cm,height=6.5cm]{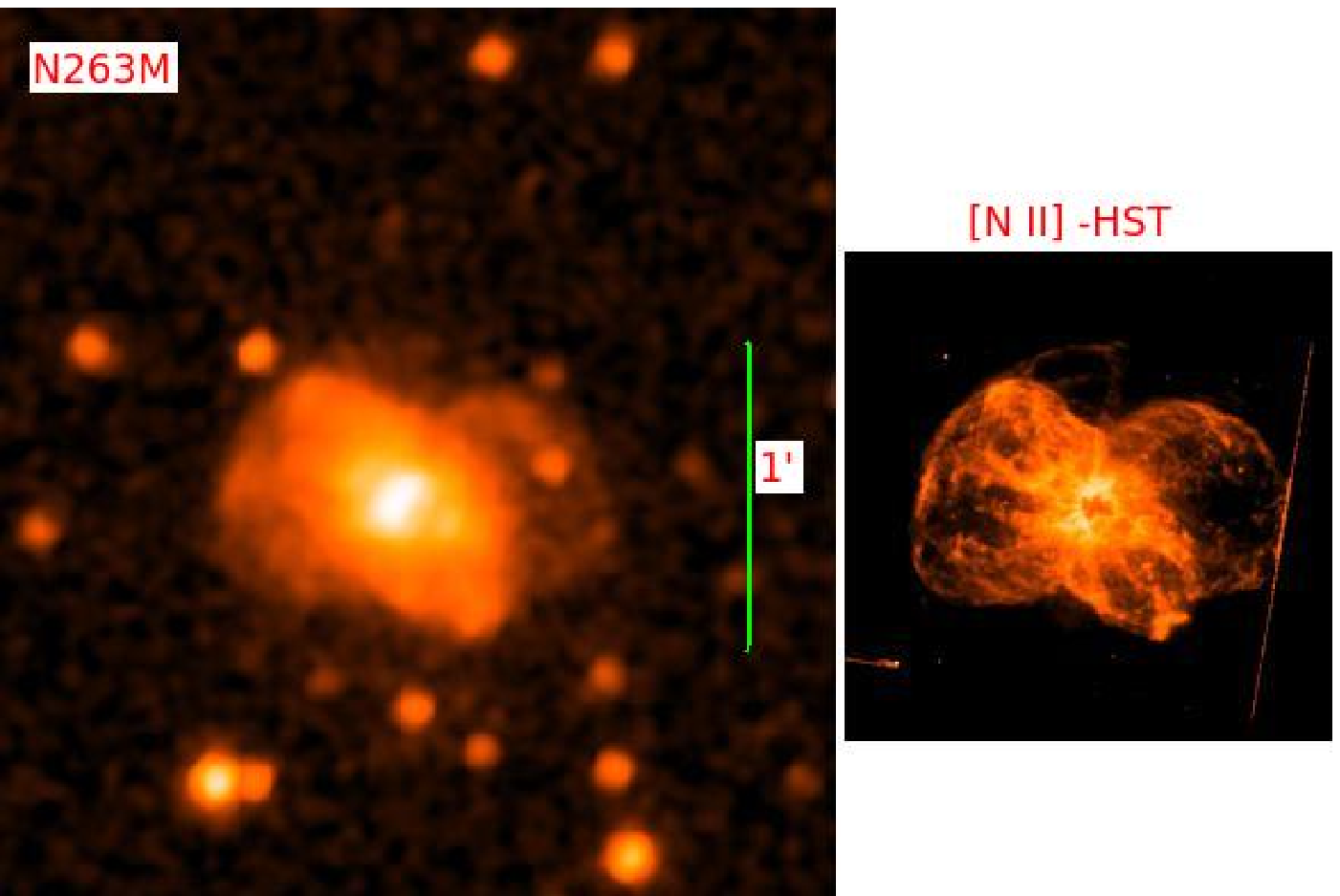}
\caption{Top left panel: Images of NGC 2440 in the filter F169M (left) is
 shown along with [O\,{\sc iii}] 5007\AA  obtained with HST.  The FUV halo
 and jet are indicated in UVIT image. In the right panel the same image is shown
 with the three bipolar orientations (PA 35$^\circ$ (a) and 85$^\circ$ (b)
 are shown.
  Bottom left panel shows UVIT  image of NGC 2440 in F172M shown along with HST
 image in [N\,{\sc ii}] 6584\AA. The bottom right panel shows the UVIT image in N265M
 compared with the HST image in [N\,{\sc ii}] 6584\AA. }
\end{figure*}

            Very detailed Hubble Space Telescope (HST) images have been
 discussed by Szyska et al. (2009) who also identified the elusive central star.
 The optical narrow band images show two main lobes with complicated clumpy small scale structure
 in the east-west direction separated by a dark lane of a very dense disc of
 gas (neutral and
 molecular) and
 dust, stretching to  north-south.  It formed into  a toroid,  that obscures the
 central star with visual extinction of about 8 magnitudes
 (Matsuura et al. 2005, Peretto et al. 2007, Szyska et al. 2009 , Wright et al. 2011) .
 Meaburn et al. (2008)
 determined the distance to the nebula as 1.17$\pm$0.14 kpc from expansion
 parallax using proper motions of features in the north-west lobe. This
 estimate seems to be consistent with measurements of  proper motions from
 Hubble images of the eastern lobe (Szyszka et al. 2011).
 From 3D photoionization modelling of the nebula Wright et al. (2011)
 derived  the properties of the central star as  hydrogen deficient
 with  T$_{\rm eff}$ of 220000 K , log $g$ of 7 ,
 L$_{\rm *}$ of 14300 L$_{\rm \odot}$and  mass of 0.73-0.82 M$_{\rm \odot}$. They
 also estimated the initial mass to be around 5.5 M$_{\rm \odot}$ .

           Extensive studies of the circumstellar torus from infrared to
 radio wavelengths ( Lester \& Dinerstein 1984, Kemper et al. 2002,
 Matsuura et al. 2005, Peretto et al. 2007, and Santander-Garcia et al. 20017)
 suggest the structure is that of a broken disc containing 2.2 M$_{\rm \odot}$ of
  dust and molecular gas expanding at 8 km s$^{-1}$, presumably ejected from the star 
 some 5000 years ago, over a duration of $\sim 2000$ years.
 The torus also obscures both the star and an ionized gas disc (detected in 6 cm
 free-free continuum) around the star. Kinematical studies of the east and
 west lobes seem to suggest that an explosive event initiated a kind of Hubble
  flow (i.e. a flow in which the velocity increases outward in proportion to its distance from the
 star) in both lobes about 2200 years back (Meaburn et al. 2008, Szyszka et al.,
 2009). The formation and flow of matter probably was directed by the torus into east - west lobes.

    Our F169M image of this nebula shows faint emission lobes that extend to about 5' arc minutes on either 
   side of the central source. Faint orthogonal jets are also present on either side of the FUV lobes through the
 central source (Figure 2-bottom). These lobes and jets are not present in either of the two NUV filters or in FUV F172M.  
 Optical and IR images of NGC 6302 show brightly emitting bipolar lobes in the east-west direction with a massive torus
 of molecular gas. Dust is seen as a dark lane in the north-south direction. FUV lobes are much more extended and oriented 
  at a position angle of 113$^\circ$. The FUV lobes and jets might be remnants of earlier (binary star) evolution, prior to 
  the dramatic explosive event that triggered the Hubble type bipolar flows about 2200 years back. The source of the FUV lobe 
  and jet emission is not known, but most likely is due to fluorescent emission from H$_{2}$ molecules. The cause of the difference 
  in orientation of optical and FUV lobes is also indeterminate, although we speculate that it could be related to the
  binary interactions.

\subsubsection{NGC 2440:}

                   A different kind of FUV halo is seen in the bipolar (multi-polar) PN
  NGC 2440  in our UVIT observations. The two prominent 
  lobes of bipolar structure prominently seen in the optical images  
  (eg. HST heritage image) are resolved in the various nebular line filter images in to at least
  two interlocking, differently oriented, bipolar structures (Lopez et al. 1998). They are oriented at position angles
  (PA) of 35$^\circ$ and 85$^\circ$, with a third one at 60$^\circ$ (Lopez et al. 1998).  These multipolar structures 
   suggest changes in the direction of sporadic mass outflows from the central object (Lopez et al. 1995, Manchado et al. 1996). 
   Many molecular emissions and outflows have have been mapped. Mapping of H$_{2}$ ($v=1-0$ S(1)) rotational transition
   shows a spiky  spherical structure of $\sim 73"$ diameters (Muthu Mariappan et al. 2007, Wang et al. 2008) with spokes emanating from the centre. 
   CO (3-2) emission closely follows the PA 35$^\circ$ bipolar axis in three clumps extending to about a radius of
   $\sim 36"$ from the central clump (Wang et al. 2008). HCN, HCO$^{+}$ emission is also detected within the nebular diameter of 
   $\sim 71"$ (Schmidt \& Ziurys 2016). Cuesta \& Phillips (2000) analysed and modeled the optical nebular line filter images. 
   Ramos-Larios \& Phillips (2009) show the Spitzer images at 3.6, 4.8, and 8 $\mu$ infra-red 
   emission extending to $\sim 80"$ diameter centered on the central source where two bright nebular 
   knots (NK and SK) separated by $\sim 6.2"$ occur and define another axis. Thus the, ionized, molecular gas and dust are 
   all confined within $\sim$80" diameter centered around the centre. Logo \& Costa (2016) 
   modeled the morphokinematical structure with two bipolar components with PA 35$^\circ$ and 85$^\circ$. The nebular 
   abundances and the central star properties have been studied recently by Miller et al. (2019).   
   
\begin{figure*}
\centering
\begin{minipage}{120mm}
\includegraphics[width=9cm,height=5.5cm]{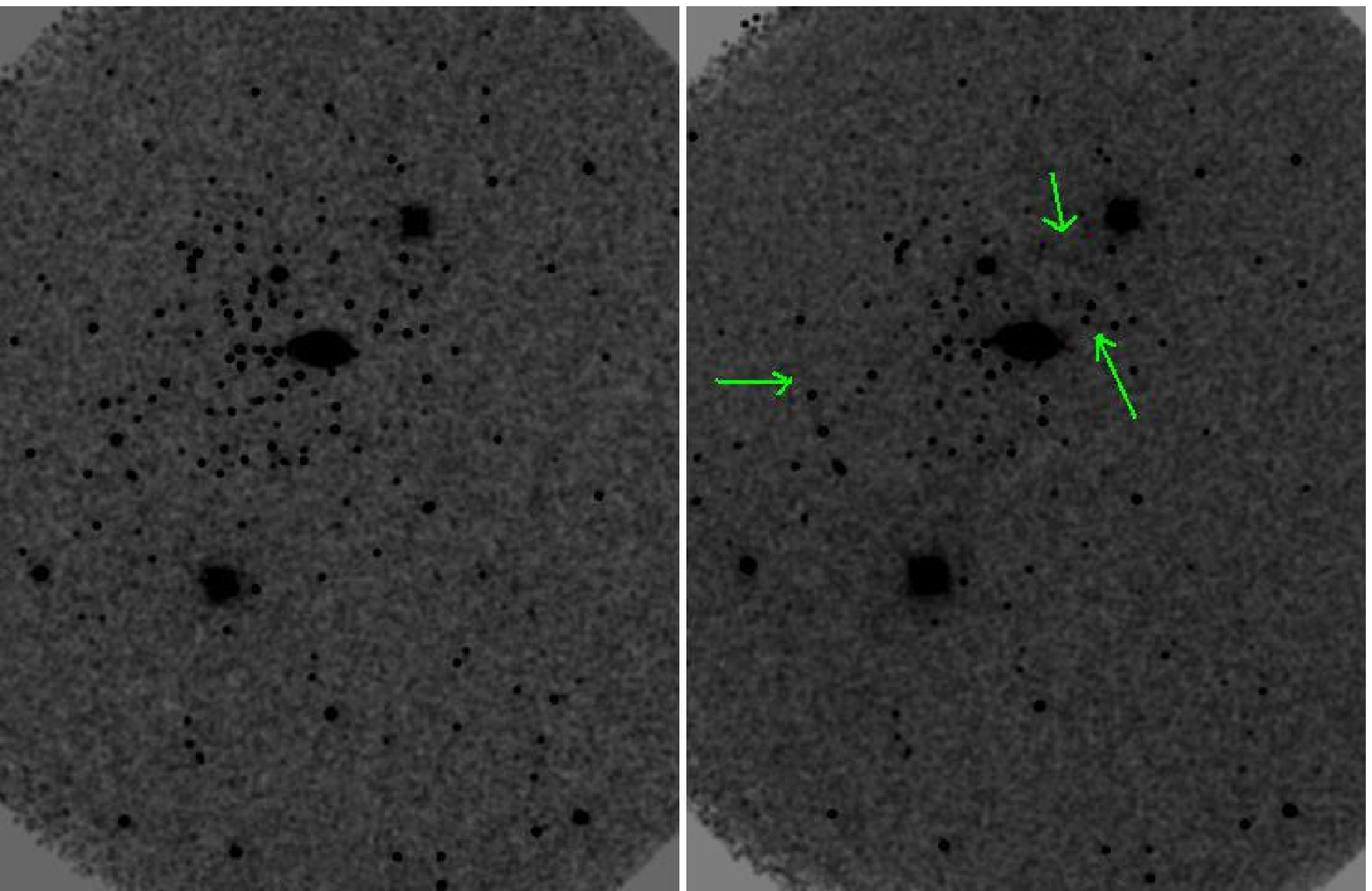}\\
\includegraphics[width=9cm,height=5.5cm]{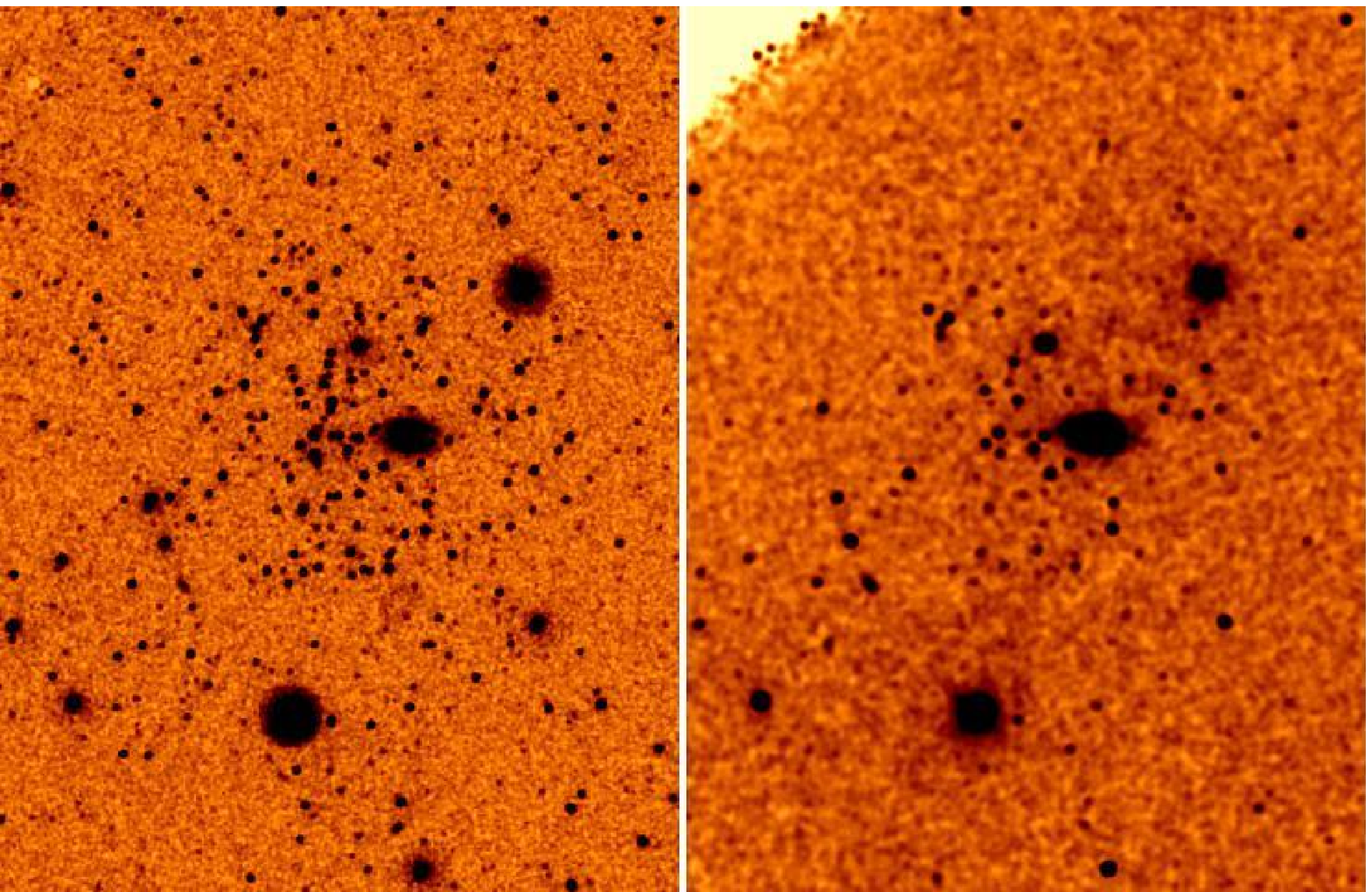}
\caption{ UVIT/FUV F154W image of NGC 2818 shows a faint, narrow, nebular,  partial
 ring like feature about 370" east of the nebula and also two nebular arcs at
 about 170" north-west of the nebula (figures on the right). These features
 are absent in  F172M (top, left) and GALEX NUV (bottom, left) images.}
\end{minipage}
\end{figure*}

\begin{figure*}
\centering
\includegraphics[width=12cm,height=9cm]{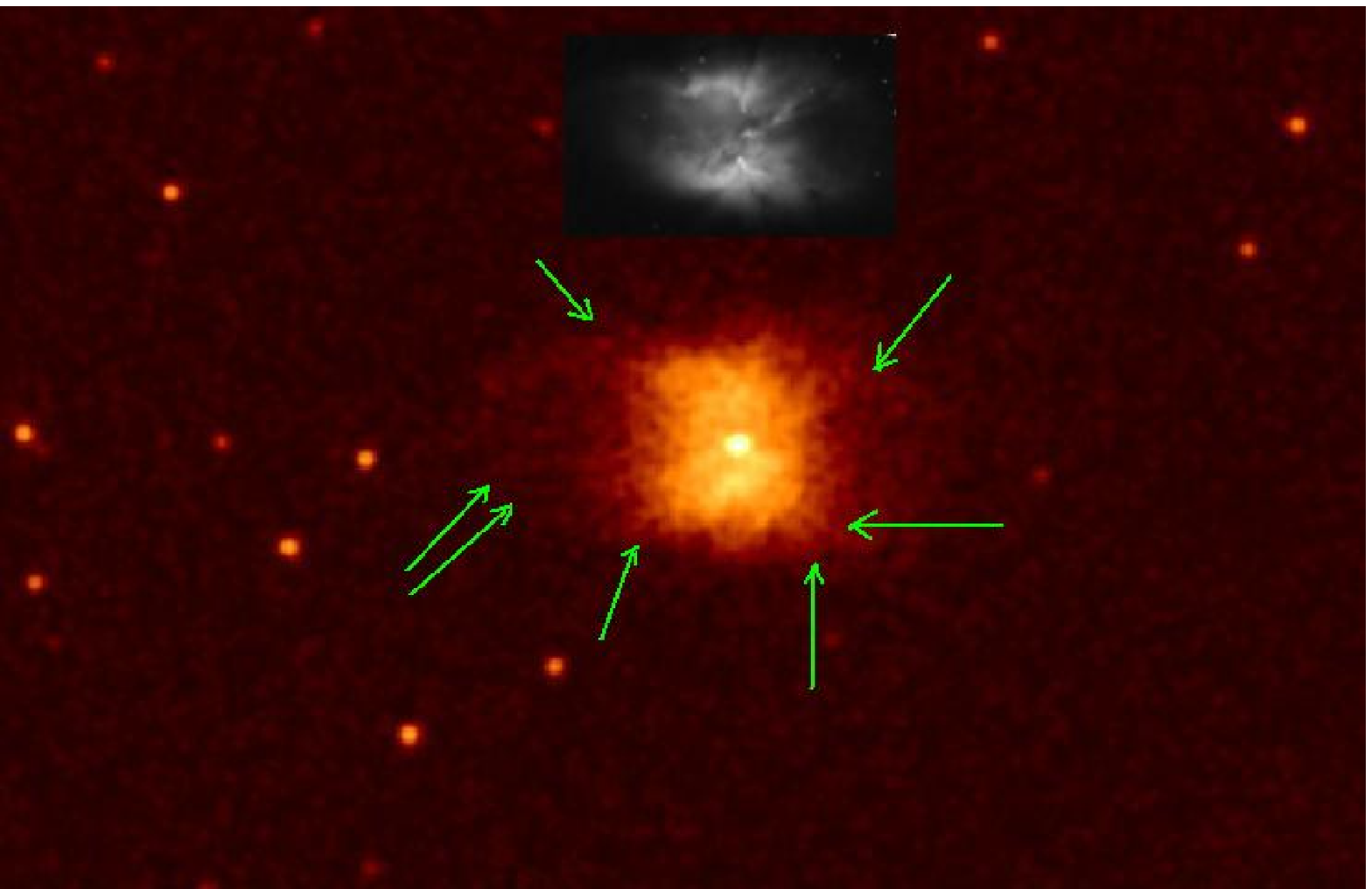}
\caption{ UVIT/FUV F169M image  of NGC 2818 is compared with [O\,{\sc iii}] 5007 \AA\, ground based,
  image (inset, Vazquez, 2012). While several features are in common, a major difference is the prominence of
 radial streamers (or emission filaments) in the F169M image, marked by green arrows. 
 The bandpass of F169M filter covers mainly He\,{\sc ii} $\lambda$1640 \AA, and C\,{\sc iv} $\lambda$ 1550 \AA\, emission lines. Presence of these streamers suggest that they  might have been swept up by strong stellar 
  wind from the central star. }
\end{figure*}

                     Our UVIT images are obtained in the FUV filters F169M and F172M as well as
   in the  NUV filters N219M, N245M and N263M (Figure 3). F169M includes the lines of  
   C\,{\sc iv} 1550 \AA, He\,{\sc ii} 1640 \AA\ close to the star whereas  F172M mostly
   displays the continuum and a weakly, the N\,{\sc iii}] 1760 \AA\ emission feature. The images of N245M
   includes  C\,{\sc ii}] 2326 \AA, N263M Mg\,{\sc ii} and the continuum. Comparison of UV
   images with optical nebular lines ground based (Lopez et al. 1998, Cuesta \& Phillips 2000)  
   and HST show that N245M and N263M are very similar to[N\,{\sc ii}] 6584 \AA where as 
   F172M image is similar to that of continuum emission (Cuesta \& Phillips 2000) both in
   size as well as in the presence of the features consistent with low excitation line
   contribution. However, the images obtained with F169M  differ from those of the other 
   UV filters and the optical lines. The bright central part of the image shows broadly 
   similar to the [O\,{\sc iii}] 5007 \AA image and shows the two bipolar systems at 
    PA  35$^\circ$ and 85$^\circ$ (Figure 4 ). The orientation of the central knots also is similar. 
               
                     The most intriguing features of the F169M image are the faint halo extending
   beyond the central bright nebula in the north-east (NE), and to a lesser extent, on the south west (SW).
   The faint halo extends beyond  the $\sim 80"$ nebular diameter estimated from 
   dust and molecular emission. It extends to  $\sim 38"$ beyond the bright nebula on the
   NE side and $\sim$18" on the SW. The axis of this halo seem to coincide with PA  35$^\circ$
   nebular axis.  In addition, the most interesting is a thin jet that extends to $\sim 44"$ to 
   the south west beyond the nebula, parallel to the PA  $35^{\circ}$ axis. (the bright 
   condensation at the southern end of the jet is a field star. It is also present in the F172M
   image).  The FUV halo in NGC 2440 is similar in nature to the other two systems NGC 40 and
   NGC 6302 and most likely caused by the fluorescent emission from cold H$_{2}$ molecules, excited 
   by the diffuse UV radiation of the hot central star.  This cold H$_{2}$ might be a product of much 
   earlier mass loss from the system when it (or possibly, the primary) was on early AGB phase. 
   The FUV jet might also be an earlier ejection from the binary system.

\subsubsection{NGC 2818:}

                      The bipolar PN NGC2818 presents a different kind of FUV emission.
  Instead of a FUV halo around the PN, NGC 2818 shows FUV emitting circum-nebular arcs
  at a distance from the nebula, possible remnants of much earlier mass ejections and
  mass-loss.

             The PN NGC 2818 is one of the few known PNs that are members 
  of  galactic clusters which makes it possible to estimate a   lower limit to
  the original  main sequence mass from the cluster turnoff. In the cse of
   NGC 2818 it is around 2.0 to 2.2 M$_{\rm \odot}$. NGC 2818 is a high excitation
  nebula with lines of He\,{\sc ii}, C\,{\sc iv}, N\,{\sc v} same time strong
   lines of low excitation as well as vibrational-rotational lines of H$_{\rm 2}$ in the near and mid-IR.
 The Spitzer images in mid-IR wavelengths show dust emission extending beyond the optical nebula,
 particularly on the western lobe (Hora 2007). From the optical spectral
 analysis the T$_{\rm eff}$ of the central star is estimated to be $\sim$ 169000 K (Mata et al. 2016). 
  Very detailed Hubble Space telescope (HST) images (Hubble Heritage image collection) have been
  discussed by Vazquez(2012)  along with its kinematical structure. The kinematical age of the wide lobes is estimated as $\sim$ 8400$\pm$3400
 years.  The optical narrow  band images show bipolar lobes  in the east-west direction with complicated
  small scale structure and  a pinched, hourglass type narrow equatorial waist
 in the middle  stretching to  north-south. The semi-major axis is estimated
 to extend to 75" through
 optical nebular lobes  in east-west and a minor axis extending
 to 55" north-south with 14" diameter central
region, which is potentially the remnant of an equatorial enhancement
 A  number of cometary knots are seen in images of low excitation lines eg.
 [N\,{\sc ii}],  that are preferentially located inside a radius of 20"
 around the central star Vazquez(2012).

\begin{figure*}
\begin{minipage}{120mm}
\includegraphics[width=9cm,height=5cm]{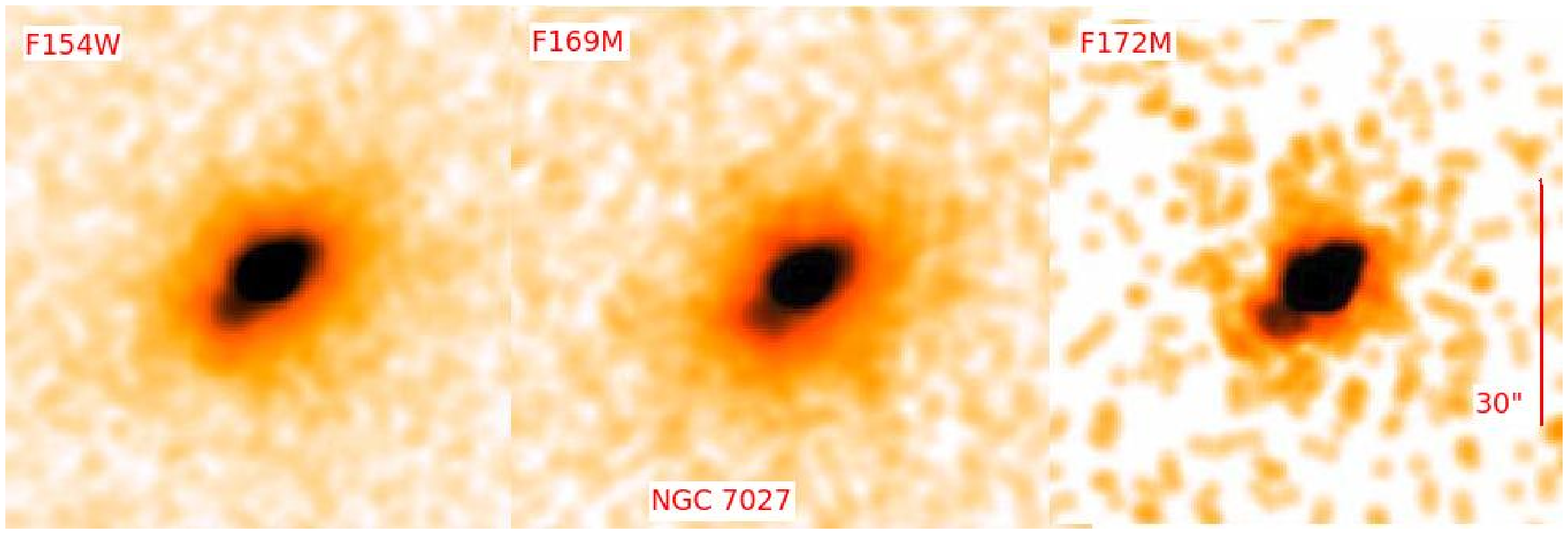}
\includegraphics[width=9cm,height=6cm]{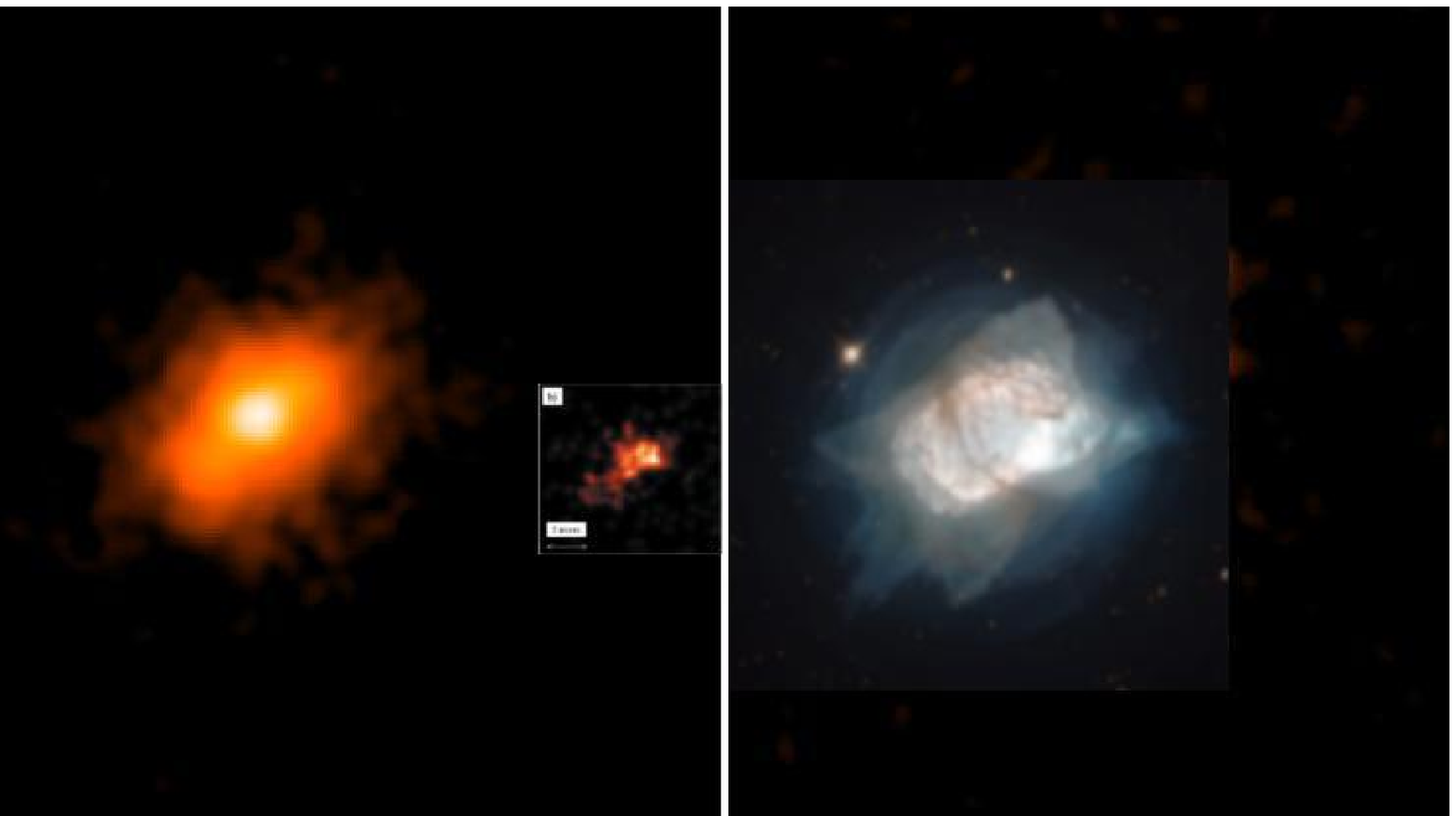}
\end{minipage}
\caption{(Top):FUV  F154W,F169M and F172M images of NGC 7027. Note that the
 south-east part is fainter and affected by extinction. FUV is very sensitive to dust
 extinction. The differential extinction can be studied from UV images.}
\caption{ (Bottom): FUV  F154W image (left) is shown along with the x-ray (Kastner et al. 2001)
 and HST optical image. The faint circular rings seen the optical image are absent
 in FUV image.}
\end{figure*}

\begin{figure*}
\centering
\begin{minipage}{120mm}
\includegraphics[width=9cm,height=4cm]{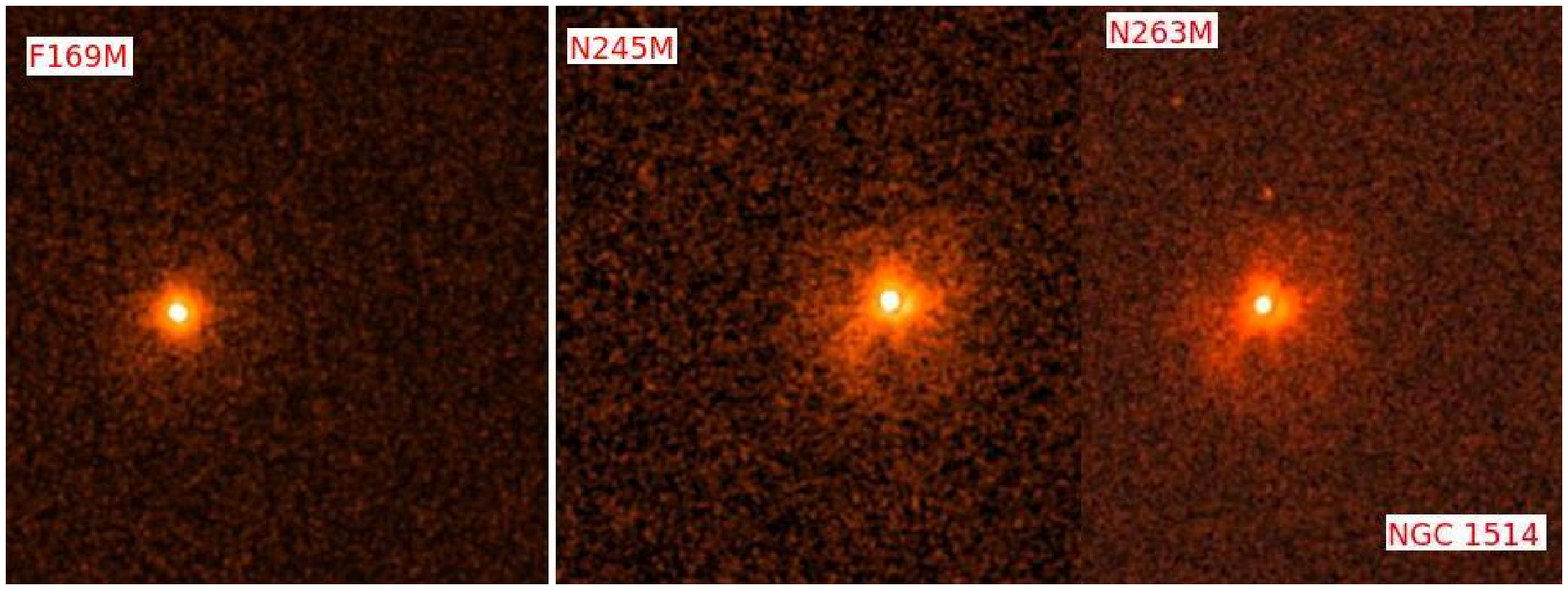}
\end{minipage}
\begin{minipage}{120mm}
\includegraphics[width=9cm,height=6cm]{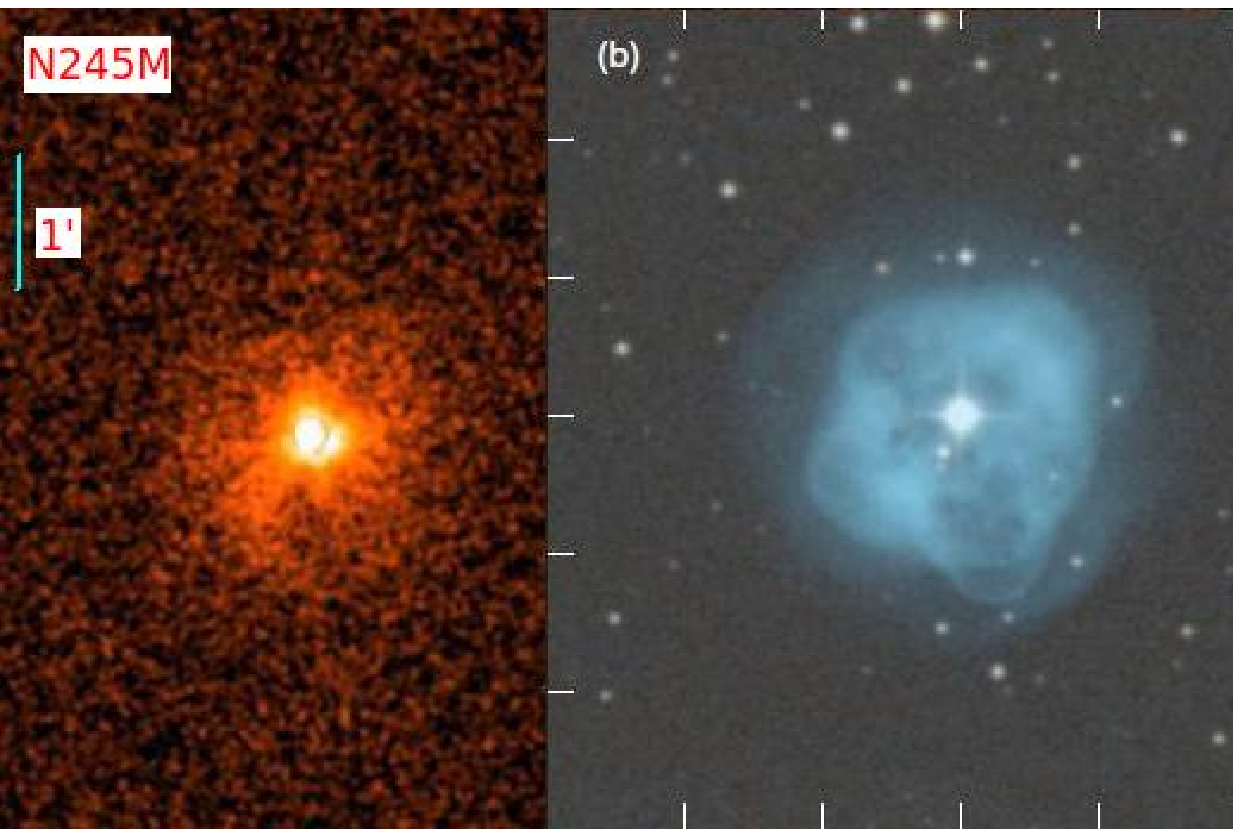}
\end{minipage}
\caption{Top: FUV UVIT F169M (left),N245M and N263M (right) images of NGC 1514.
 The nebula is brighter in NUV N245M than in FUV F169M. Bottom: The UVIT N245M
 image of NGC 1514 is compared with optical image (bottom right) (Resseler et al. 2010). The UV
 emission is confined to the inner ring and centered around the central star. }
\end{figure*}

    The bipolar  planetary nebula, NGC 2818 and the open cluster 
   have been imaged in three
 far-ultraviolet (FUV) filters , F154W: $\lambda$$_{\rm eff}$ of 1541,
 F169M :  $\lambda$$_{\rm eff}$ of 1608  and F172M:
$\lambda$$_{\rm eff}$ of 1717  with UVIT.
 The F154W image shows faint emission of a partial  nebular ring  and
 couple of nebular arcs (shell) that surround the central nebula at a
  distance of  370" and 170" from the central star (Figure 5). F169M image
 also shows traces of
  these features but not as prominently as in F154W image. But the images in
   F172M filter,
   NUV from GALEX and optical filters  do not show any trace of these emission
 features. The FUV emission from partial ring at a distance of
  6.4 pc from the star suggests an ejection that took place about
  60000 years back (or more) from the central star. The observed  expansion velocity of
  105 km s$^{-1}$ of the polar
 lobes  and the distance of 3.56 kpc  determined from Gaia
 parallaxes for both the cluster and the nebula suggests such an age. This is
 by far the most distant and oldest relic of mass ejection observed for a
  planetary nebula. The FUV emission in these nebular features is most likely
  due to UV  fluorescent H$_{2}$ molecules. From the T$_{\rm eff}$ and luminosity 
  of the star it appears that  enough stellar UV   radiation
  reaches the nebular arc to produce sufficient H$_{\rm 2}$ fluorescence. The original
  formation of the shell (or a ring) might have involved shocks or high temperature and
  high velocity gas, but in due course that gas  has recombined  and cooled  to the
  present cold  molecular gas.

  The FUV images of the central
  bipolar nebula show bright
  emission region dominated by  He\,{\sc ii} $\lambda$1640  and to lesser extent    
  C\,{\sc iv} $\lambda$ 1550 emission,
  around  the star. Another prominent morphological aspect of FUV emission, particularly seen in
  F169M image is the presence of radial filaments (Figure 6)
  diverging from the
  central star in almost all directions. These filaments are spread more in the
  direction of eastern lobe than towards western lobe. They   extend to about 48"
  into eastern lobe. These filaments are  much more prominent in F169M image than in
   [O\,{\sc iii}]  or F172M  images,
  suggesting that they represent He\,{\sc ii}  ( and C\,{\sc iv} ) line
   emitting  regions. The filaments have a
  width of about  0.065 pc at the distance of the
  nebula.   This structure of the filaments  is very
  similar to the radial rays surrounding the main ring in Helix nebula as shown
  by  O'Dell (2004) . These radial filaments  seem to provide channels for the 
 hot stellar wind to flow. 
 
                 It is amazing that FUV studies could bring out relics of 60000 years
  past mass-loss from the pre-CSPN star of NGC 2818. 
          

\begin{figure*}
\begin{minipage}{120mm}
\includegraphics[width=17cm,height=4cm]{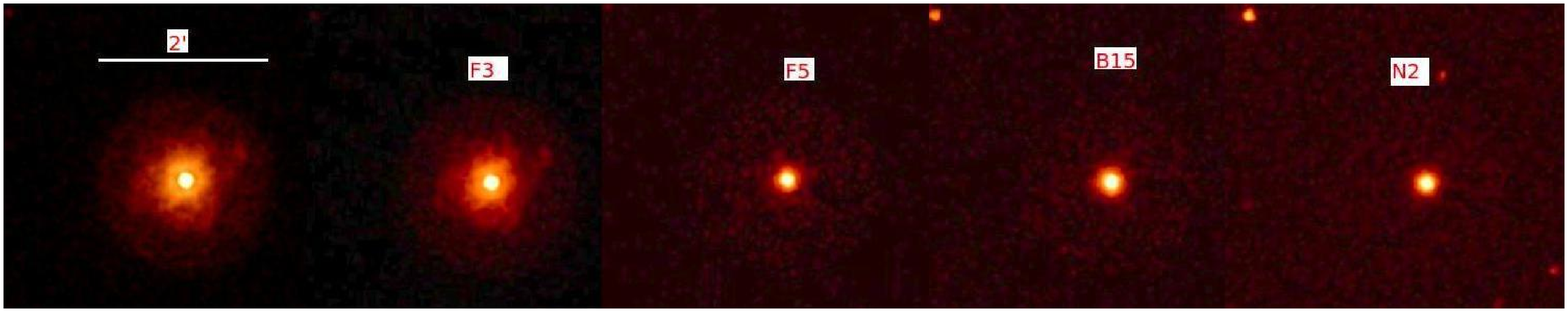}
\end{minipage}\\
\begin{minipage}{120mm}
\includegraphics[width=9cm,height=6cm]{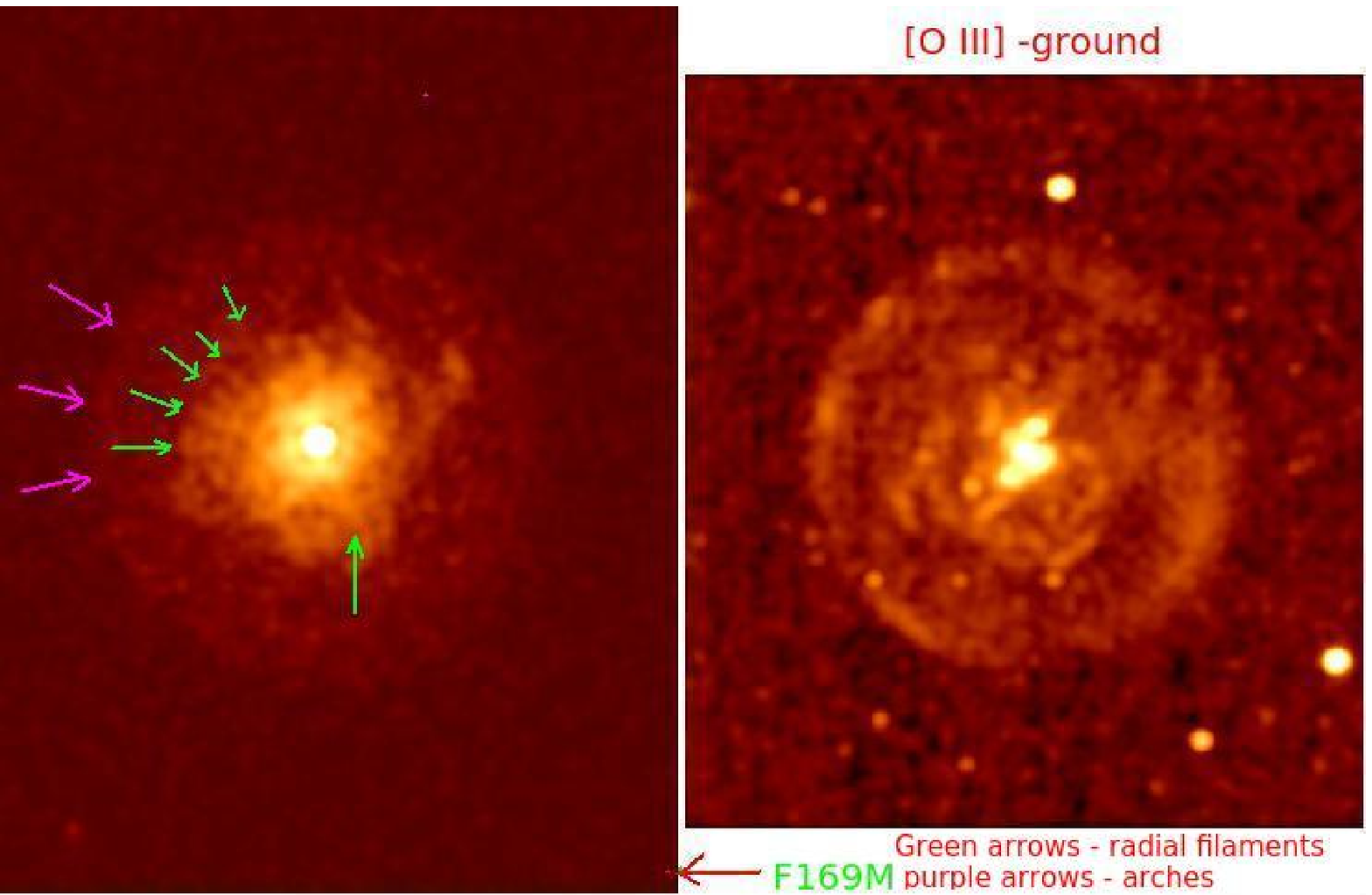}
\end{minipage}
\caption{(Top:) PN A30 in various UVIT FUV and NUV filters is shown in the top panel. 
         (Bottom:) The UVIT/FUV F169M image of PN A30 (left) is compared with ground based  [O\,{\sc iii}] 
image (right). The Knots and cometary-tail like filaments in the centre of the [O III] image appear to have
counterparts in the F169M image. The [O\,{\sc iii}] image was supplied by Arturo Manchado (personal comm.).}
\end{figure*} 

\subsubsection{NGC 7027:}

                       Although our analysis of this PN is on going, we illustrate 
 here F154W, F169M and F172M filters image (Figure 7) from our observations and point out some
   interesting features.  NGC 7027 (PN G084.9-03.4) -- also known as the "Magic carpet" nebula  or "Pink pillow" nebula -- 
is  located at 1 kpc (Zijlstra et al. 2008) and has  a kinematical age of just 600 years (based on its radio flux -
Masson 1989). It is a  compact and young PN, one of the brightest nebulae in the sky and the most extensively studied one.  NGC 7027 is
a carbon-rich nebula with a very high-excitation spectrum showing  lines of  O\,{\sc iv}, Mg\,{\sc V} etc.  It hosts one of the
hottest central stars known to date, with a T$_{\rm eff}$ $\sim 200000$ K (Latter et al. 2000). A small, essentially ellipsoidal, expanding
ionized shell surrounds the central star (Masson 1989). Further outwards, a thin  shell indicates the presence of H$_{2}$
a photo-dissociation region (PDR) and shows signs of recent interaction with collimated outflows (Cox et al. 2002). The nebula
also shows many molecular emission lines, e.g. lines of CO, CH$^{+}$, H$_{2}$O, and even HeH$^+$ ion have been detected from beyond the 
PDR (Wesson et. al 2010, Santander-García et al. 2012, Gusten et ai. 2019).

\begin{figure*}
\centering
\begin{minipage}{120mm}
\includegraphics[width=9cm,height=6cm]{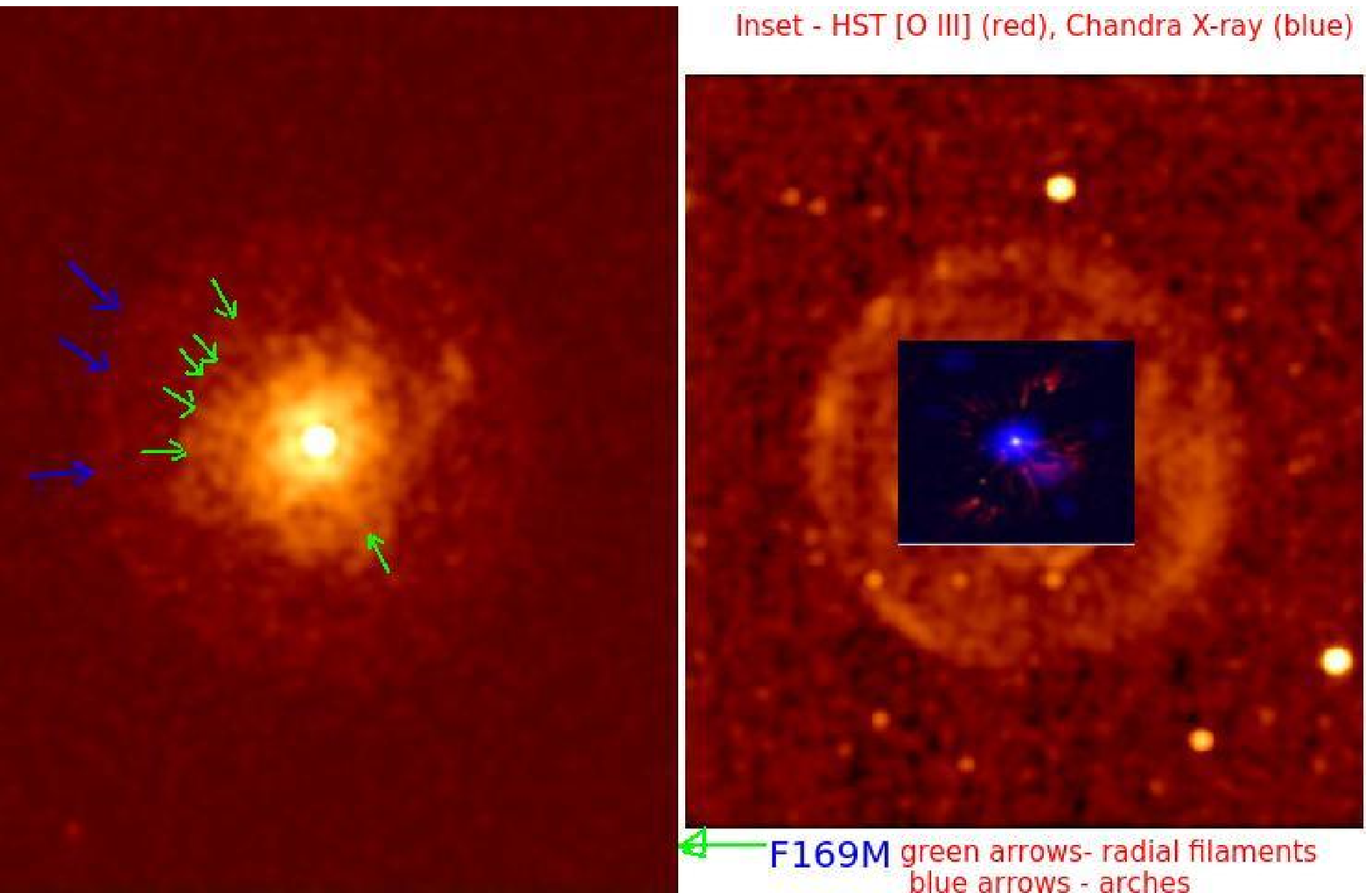}
\end{minipage}\\
\begin{minipage}{120mm}
\includegraphics[width=9cm,height=8cm]{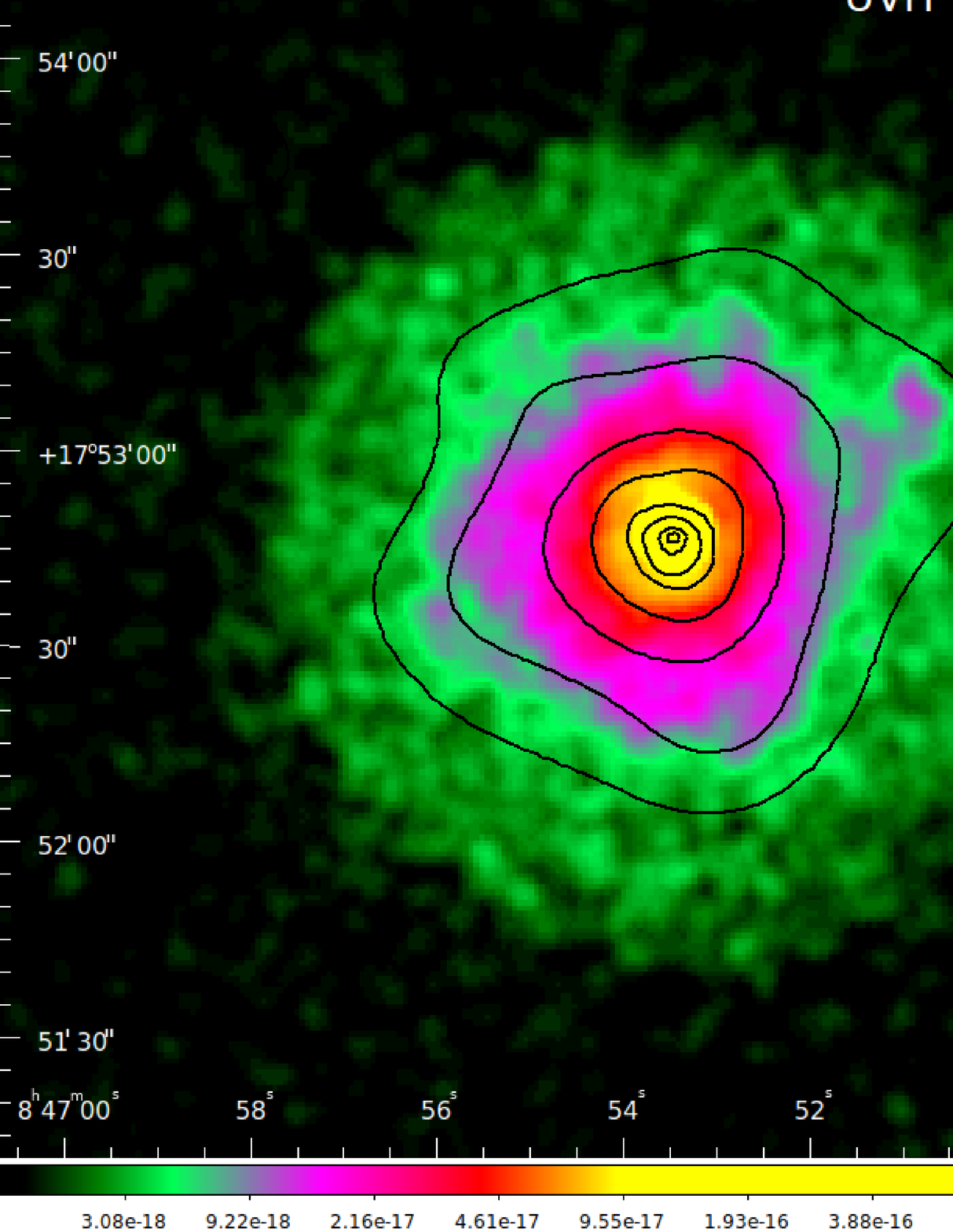}
\end{minipage}
\caption{(Top, left): A30 image in F169M showing radial streamers from CSPN showing
  channels of streaming stellar wind (green arrows). The pink arrows point to
  the arches where the channeled flow hit the outer boundary. (Right):
  Ground based [O\,{\sc iii}] image superposed by an inset showing the region of Chandra X-ray
  emission (Guerrero et al. 2012). (Bottom): Image of A30  F169M superposed by the XMMNewton X-ray continuum
  counters.  }
\end{figure*}

\begin{figure*}
\includegraphics[width=17cm,height=7cm]{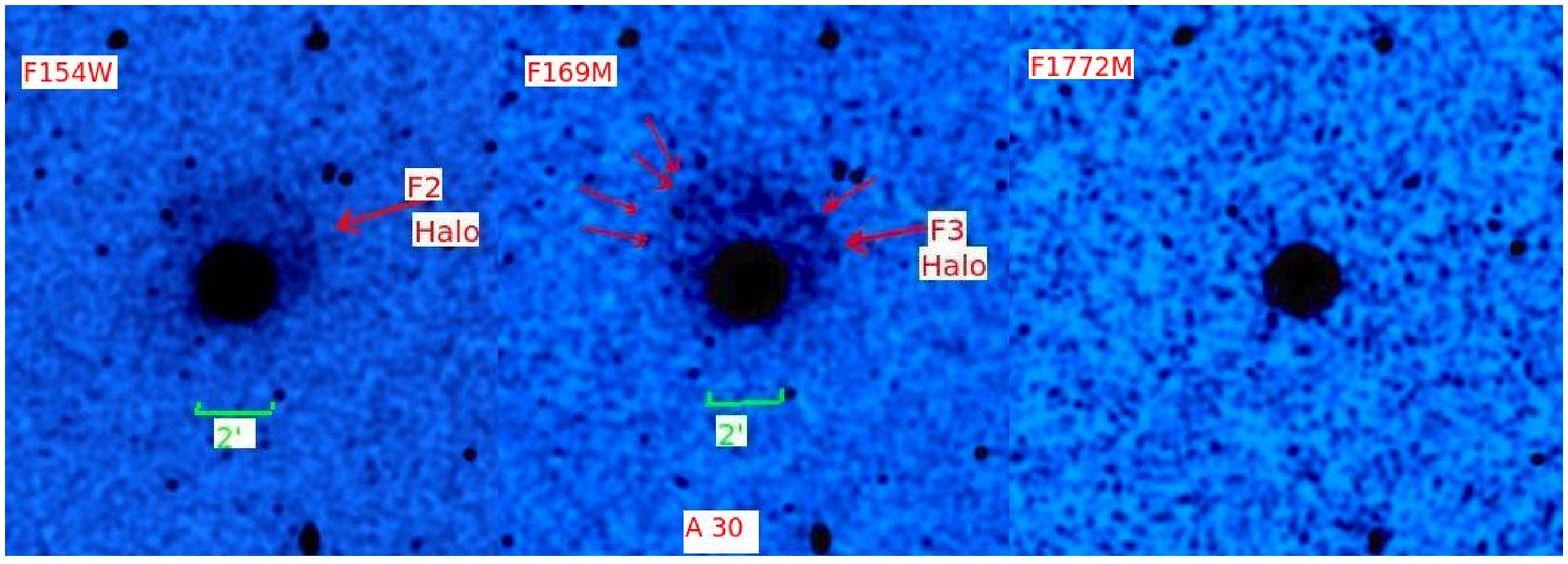}
\caption{ Images of A30  in F154W , F169M and F172M showing the FUV halo around F154 and F169M
  images and the absence of it around F172M image. }
\end{figure*}

The PN was discovered
to be an X-ray source by Kastner et al. (2001), who attributed the
X-ray emission to shock heating by a fast wind from the central
star impacting the slow wind which the progenitor star ejected
while on the asymptotic giant branch (AGB).
Zhang et al. detected  in the spectrum of this bright young PN of
Raman-scattered O\,{\sc vi} features at 6830\AA\ and 7088\AA\, pointing to the
 existence of abundant neutral hydrogen around the ionized region

      FUV is very sensitive to the internal and external extinction.
 NGC 7027 has high and variable extinction across the nebula that has
 been mapped by Walton et al. (1988). Montez et al. (2018) link the 
 X-ray emission to the distribution of extinction across the nebula.
 UVIT images in FUV would reflect such extinction variations (Figure 8) that
 would explored later. UVIT images do not show presence of FUV halo
  or  arcs as in other bipolar PNs listed earlier although deep
 images in optical (HST) do show faint circular rings around the 
 main nebula. One possible reason is that because of high internal
 extinction no UV photons from the central star reach to the outer
 H$_{2}$ region.

\subsection{Round Nebulae -FUV Halos }

            The members of this group that were observed by UVIT are
 NGC 1514, A 30 , EGB 6, and NGC 3587

\subsubsection{NGC 1514:-shining fluid!}

          Observing NGC 1514 on 1790 November 13, William Herschel  called
  it `a most singular phenomenon!'.This PN is unusual because of its very
 bright central star and large diameter low surface brightness nebulosity.
 The observation by Herschel is termed as `an important event in the
 history of astronomy' (Seaton 1980) because of his realisation that `we
 therefore either have a central body which is not a star or have a star
 which is involved in a shining fluid, of a nature totally unknown to us'
 (Seaton 1980). It turns out that there is not one star at the centre but
  two, a sdO + A0III. Recent observations show that the binary system has a  
 period of 3306 days and eccentricity of 0.46. The estimates of the mass for
  the cooler secondary is about 2.3 M$_{\rm \odot}$ and the hot primary 
 is of 0.9 M$_{\rm \odot}$ (Jones et al. 2017).                      

    Ressler et al. (2010) described the morphological structure of NGC 1514
   (see their figure 1) as a nebula
   with two shells, inner and outer with diameters of 132" and 183". The
   inner shell has numerous bubble structures at its edge pushing 
  into the outer shell.
    The monochromatic optical images of NGC 1514  show that
   the amorphous appearance of the nebula contains very little
  nebular emission within about 30" of the nucleus (Balick 1987). This is
    also confirmed by the absence of C\,{\sc iii} ] $\lambda$1909
  emission  which is normally the strongest nebular feature in IUE spectra
  of PNe (Ressler et al. 2010). 
     
      A pair of infrared  bright  axisymmetric rings that surround the
 visible nebula were discovered by Ressler et al. (2010), particularly dominant
  in 22 $\mu$ m.
   Such a structure is not suggested in any of the visible wavelength images
  which is probably resulted from binary interaction.
   NGC 1514 is also a X-ray source (Tarafdar \& Apparao 1988, Montez et al. 2015).

           We have obtained UVIT images in F154W, F169M, F172M as well as
  in N245M, N263M, and N279N. Although our analysis is ongoing and not
  complete ,we present
  few images and show the comparison with the optical image (a combination of
  B, V, R+I -Resseler et al. 2010) in Figure 9. The UV emission seem to 
  be mostly
  to the inner shell. There is UV (nebular)  emission within 30".
  The nebular extent in FUV is less than that in  near UV and optical. There
  seems to be streams connecting the central region to the inner shell
  particularly in F245M. The structure in NUV F245M is very similar to the
  optical except the bubbles look sharper.

\subsubsection{ Born-again Planetary Nebula A30 -}

                Abell 30 (PNG208.5+33.2, A30) is archetypal born-again PN of about 2 arc
  minute diameter. The central star of A30 is believed to have undergone a very late
 thermal pulse (VLTP)   that
 caused the ejection of hydrogen deficient material, prominently
 seen in the light of [O\,{\sc iii}] lines, about 850 years back. The inner parts
 of the
 nebula are filled with this material whereas the outer rim of the nebula
 is of H-normal composition and of about 12500 yrs of age. A30 is also an X-ray source
 showing both a diffuse source covering the inner few arc
  seconds covering the hydrogen deficient knots and a point source located on the
 central star.


\begin{figure*}
\includegraphics[width=17cm,height=7cm]{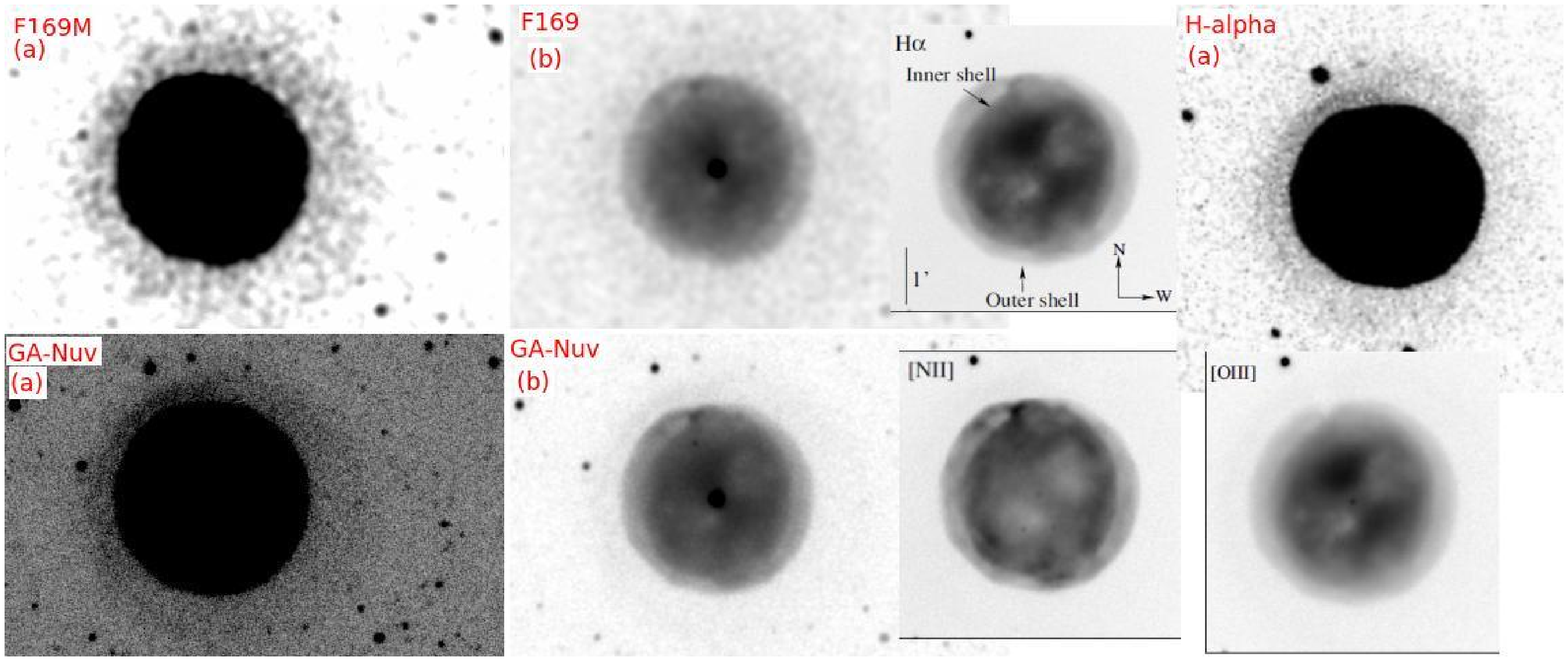}
\includegraphics[width=12cm,height=7cm]{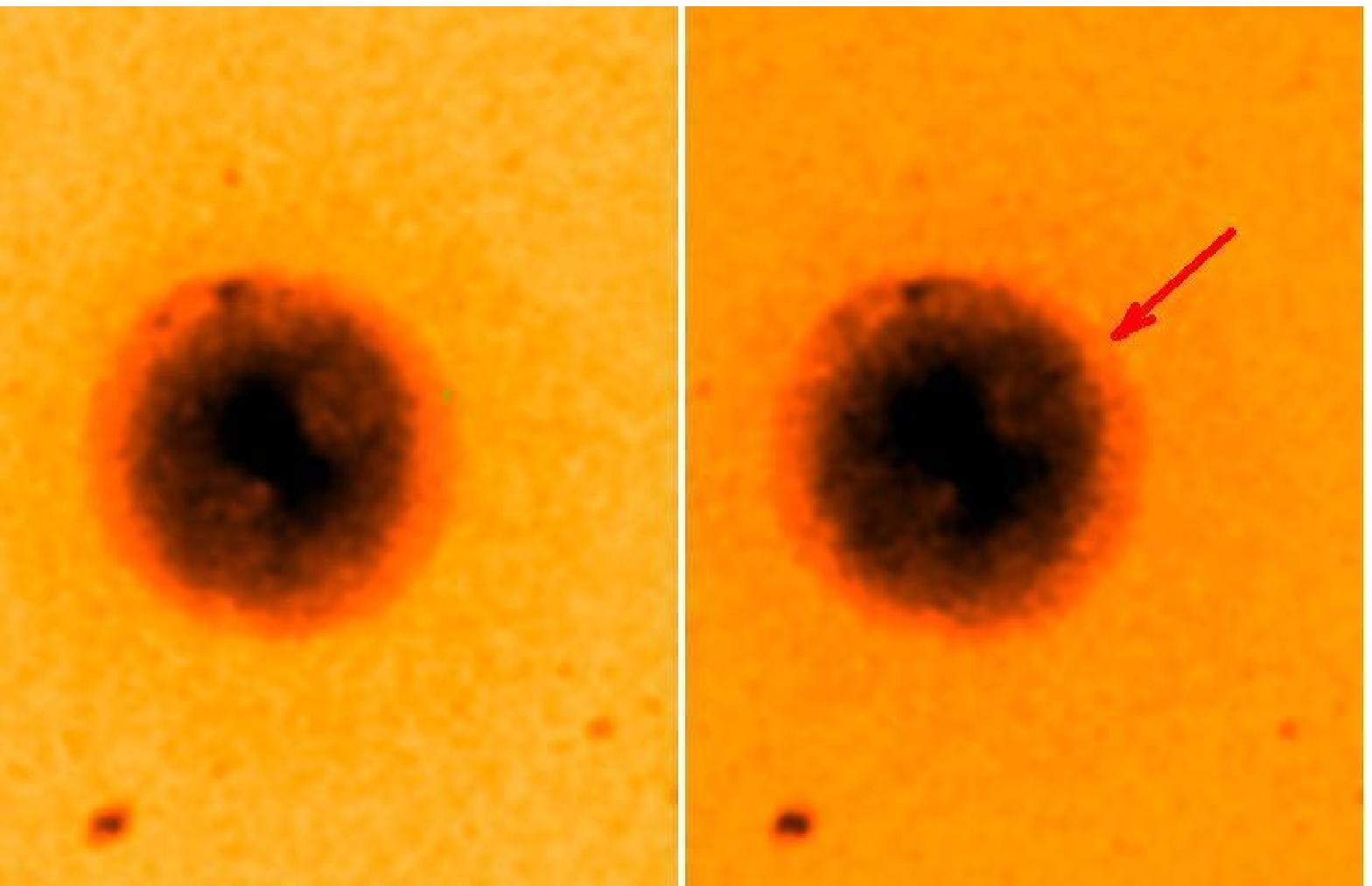}
\caption{ Top: Images of NGC 3587 (Owl Nebula) in F169M and Galex NUV showing the halos
  around the deep images (marked as (a)). The nebula in F169M and Galex NUV with two shells
  are shown marked as (b). Ground based H$\alpha$, [N\,{\sc ii}] and [O\,{\sc iii}] images
  (from Garc\`{i}a-D\`{i}az et al. 2018) are shown for comparison with UV images. The features of
  F169M and Galex NUV are similar to [O\,{\sc iii}] and H$\alpha$ respectively.
  Bottom: Image of NGC 3587  in F154W (left) and F169M (right). F169M image shows a jet like
  feature (shown by the  arrow) that is absent in F154W image. }
\end{figure*}

           Born-again planetary nebulae (PNe) are believed to have
experienced a VLTP (Iben et al. 1983)
while the star was descending the white dwarf cooling track.
During this event, the remaining stellar helium envelope reaches
the critical mass required to ignite its fusion into carbon and oxygen
(e.g., Herwig 2005; Miller Bertolami \& Althaus 2006; Lawlor
\& MacDonald 2006); the sudden increase of pressure leads to
the ejection of the newly processed material and, as the stellar
envelope expands, its temperature decreases and the star returns
in the Hertzsprung–Russell (HR) diagram to the locus of the
asymptotic giant branch (AGB) stars. Soon after, the contraction
of the envelope will increase the stellar effective temperature,
boosting the UV flux, and giving rise to a new fast stellar wind.
 So far, the only bona fide born-again PNe are Abell 30 (A 30), Abell 58 (A 58, Nova Aql 1919), 
Abell 78 (A 78), and V4334 Sgr (the Sakurai’s object).
Among them, A30 and A78 are the more evolved ones, with
large limb-brightened, H-rich outer nebulae surrounding the
H-poor, irregular-shaped structures that harbor the “cometary”
knots in the innermost regions (Jacoby 1979; Hazard et al. 1980, Meaburn \& L{\'o}pez
1996; Meaburn et al. 1998).  HST images in the [O\,{\sc iii}] emission line of the 
central regions have revealed equatorial rings (ERs) and compact polar outflows
(POs) in the central regions of both PNe (Borkowski et al. 1993, 1995). 
The dynamics are revealing: while the outer nebulae show shell-like structures expanding at 
velocities of 30 to 40 km $^{-1}$, the H-poor clumps present complex structures, with velocity
spikes of 200 km $^{-1}$ (Meaburn \& L{\'o}pez 1996; Chu et al. 1997;
Meaburn et al. 1998). The morphology and kinematics of the H-poor knots unveil
rich dynamical processes in the nebulae. The material photo-evaporated
from the knots by the stellar radiation is swept up downstream by the 
fast stellar wind, which is otherwise mass loaded and slowed down (Pittard et al. 2005). 
The interactions are complex, resulting in sophisticated velocity structures, as well as 
x-ray emitting hot gas (Chu $\&$ Ho 1995; Guerrero et al. 2012; Toal{\`a} et al. 2014).

\begin{figure*}
\includegraphics[width=15cm,height=5cm]{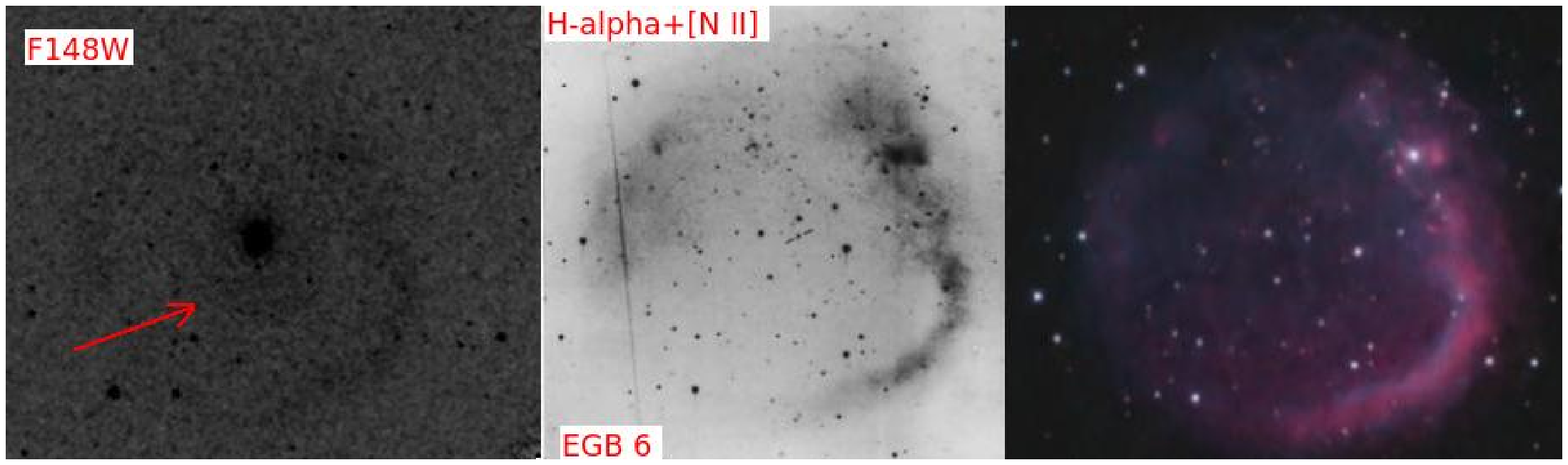}
\includegraphics[width=12cm,height=6cm]{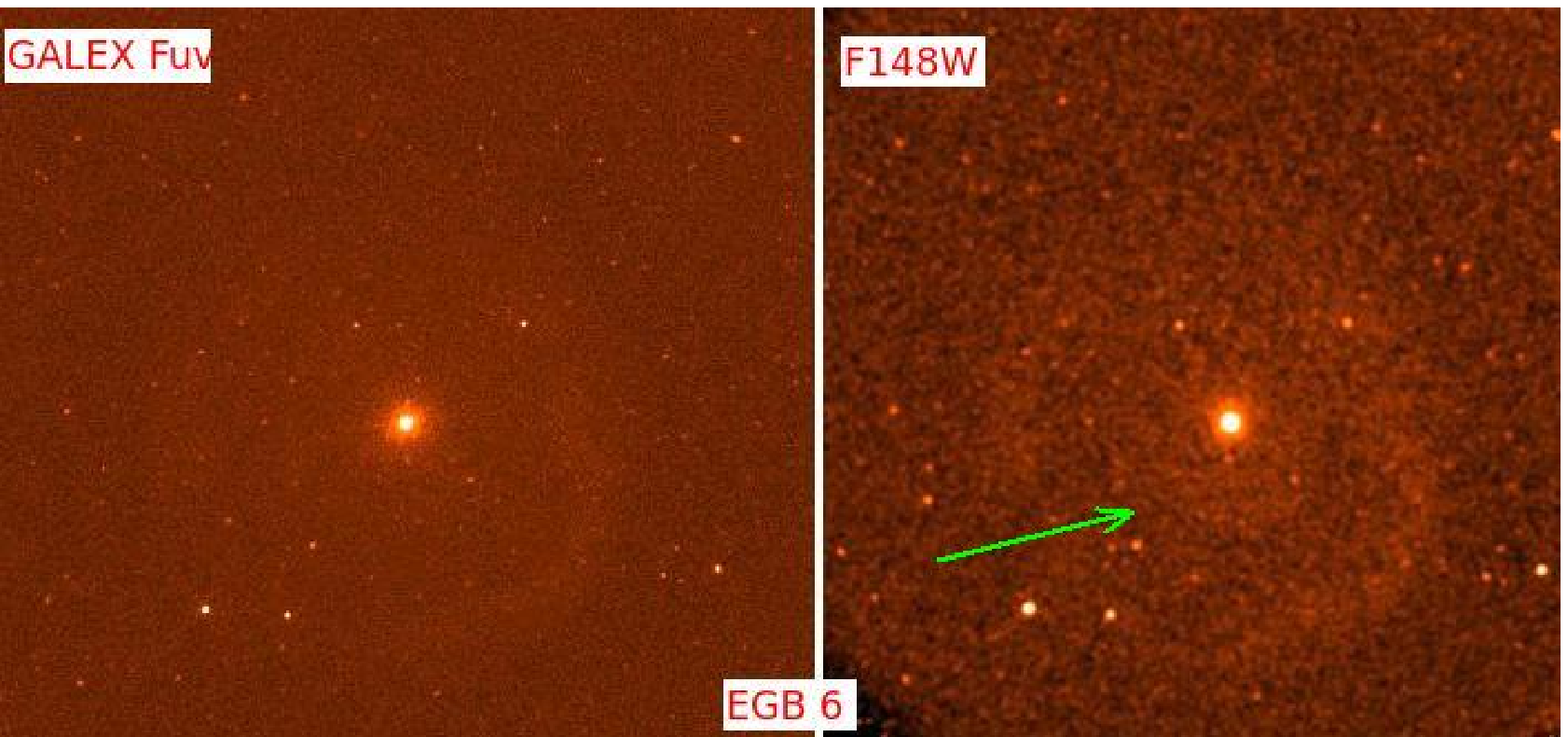}
\caption{ Top, 3 Panels :UVIT image of EGB 6 in F148W filter showing
the fainter outer nebula and smaller central nebula (shown by the arrow) along with
 an optical image
of EGB 6 in H$\alpha$ and [N\,{\sc ii}] (Jacoby et al. 1995) and an optical colour
image of the outer nebula from Don Goldman (astrodonimaging.com). Note that the inner smaller neb is not present in
the two optical images.
Bottom: Galex FUV image of EGB 6 (left) is compared with UVIT image in F148W.
 The inner small nebula is shown by the arrow.
 Note the absence of inner small nebula in the Galex image.  }
\end{figure*}


\begin{figure*}
\begin{minipage}{120mm}
\includegraphics[width=12cm,height=5cm]{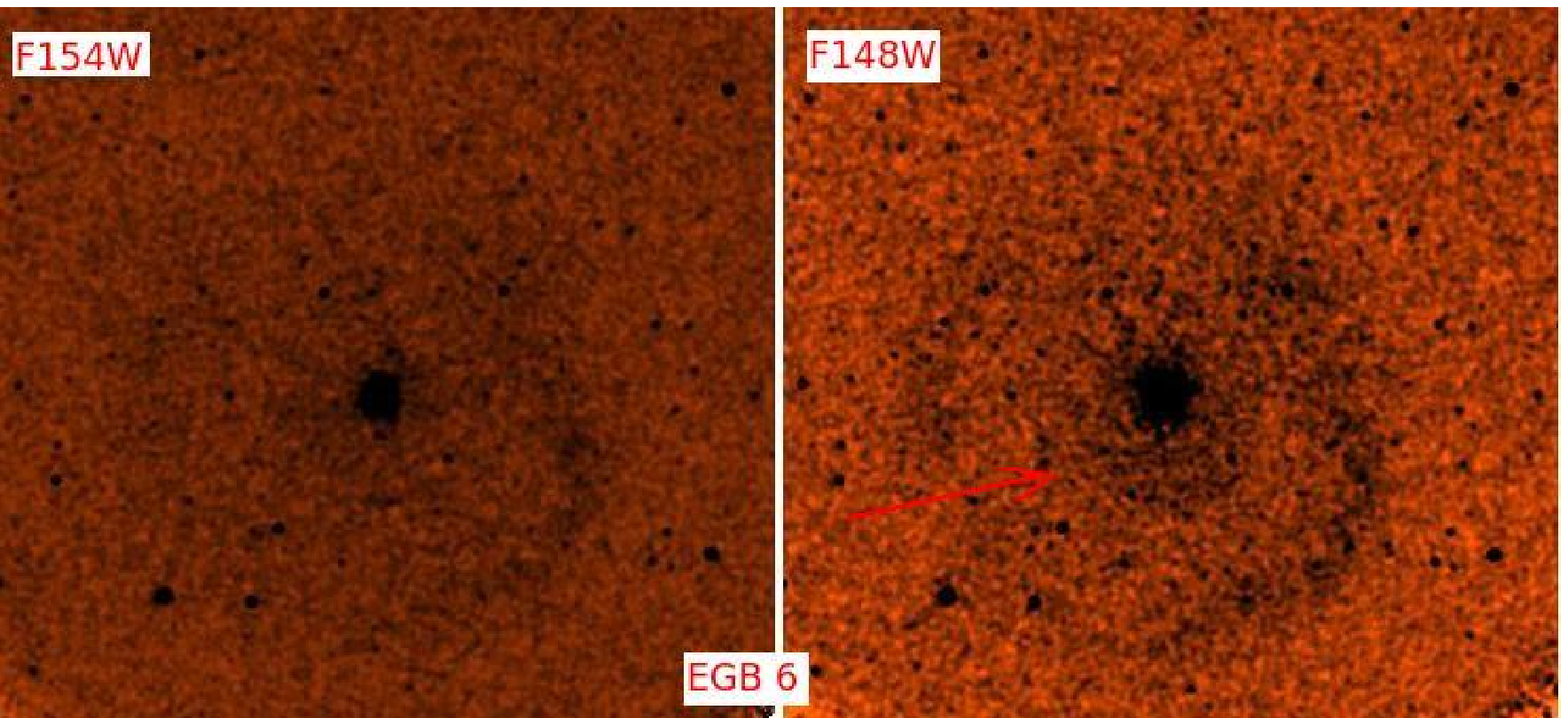}
\end{minipage} \\
\begin{minipage}{100mm}
\includegraphics[width=10cm,height=10cm]{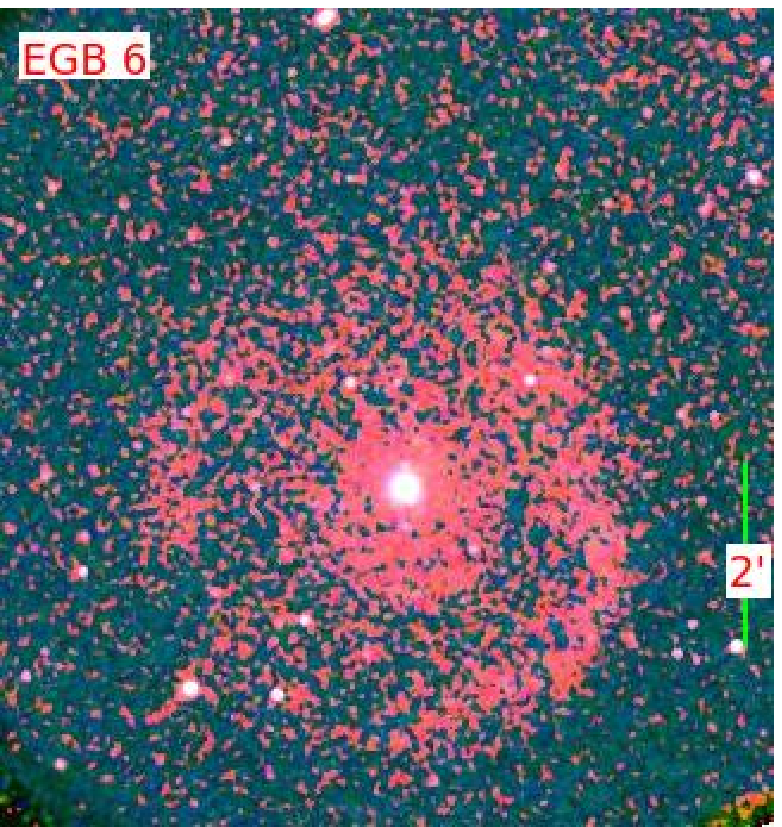}
\end{minipage}
\caption{ Top, 2 Panels: UVIT image of EGB 6 in F154W filter showinging
the fainte outer nebula and smaller central nebula along with F148W image of the outer nebula. Note that an inner, smaller,
nebula is  present in the two FUV images.
Bottom:  FUV image of EGB 6  in F148W. Note that the bow of inner small nebula is towards south-west  direction.  }
\end{figure*}

         To study the correspondence of UV emission  with the X-ray emission 
  as well as the hydrogen deficient ejecta, we imaged A30 with UVIT 
  in 3 FUV and 2 NUV filters. Two FUV filters, F154M (F2)  and F169M (F3; $\lambda_{\rm eff}$ 1608 \AA) 
transmit the high excitation lines of He\,{\sc ii}, C\,{\sc iv} etc. as shown in fig. 1).  The other FUV filter F172M 
(F5 with $\lambda_{\rm eff}$ 1717 \AA) allows mostly the nebular continuum. The NUV filters N219M (B15 with $\lambda_{\rm eff}$  
2196 \AA) and N279N2 (N2 with $\lambda_{\rm eff}$ 2796 \AA) allow mostly low excitation lines or continuum. The images of the 
nebula in these filters are shown in Figure 10. The nebula is most intense in F169M and F154W where He\,{\sc ii} line
emission dominates.  The UVIT provides a spatial resolution of $\sim 1."3$. In the present work,
 we contrast the UV images with both X-ray contours as well as ground based [O\,{\sc iii}] and H-alpha images
 (Arturo Manchado -personal communication). In the FUV F2, F3 images,  the hydrogen deficient nebular knots are not as
conspicuous as in the [O\,{\sc iii}] image (Figure 10). The FUV F2, F3 show radial streamers, which are quite 
prominent almost extending from central region (Figure 11) to the edge of the nebula. They
 provide the channels for  the material photoevaporated from the knots by the stellar
 radiation is swept up downstream by the fast stellar wind, which is otherwise mass
 loaded and slowed down (Pittard et al. 2005). At the edge of the channel where it interacts 
 with the nebular boundary, arch type structures are seen suggesting that the boundary
  is being pushed out by the flow of stellar wind swept material. 
 X-ray emission has been detected within A30 (e.g., Guerrero et al. 2012; Kastner et al. 2012; Montez et al. 2015).
This born-again PN has been studied with ROSAT (PSPC and
HRI), Chandra, and XMM-Newton X-ray satellites (Chu \& Ho
1995; Chu et al. 1997; Guerrero et al. 2012). Its X-ray emission
originates from the CSPN, but there is also diffuse emission
spatially coincident with the cloverleaf-shaped H-poor structure
detected in [O\,{\sc iii}]. The X-ray emission from both the CSPN and
the diffuse extended emission is extremely soft. The XMM/Newton x-ray
 continuum contours (Chu et al. 1997) are shown in Figure 11 superposed on F169M image.
 They are confined
 mostly to the inner nebular region where the He\,{\sc ii} dominated gas is present.
  The Chandra X-ray region is displayed in Figure 11 as an inset displaying the HST image
 of the inner 10" radius of the nebula. The X-ray region contains both diffuse emission and 
  knots.   
    
 \subsubsection {FUV Halo:}
   
            The most {\bf surprising result} of our UV imaging of A 30  is the
presence of a FUV halo in the F154M and F169M filters, extending beyond the known
 optical and NUV nebular size (Figure 12). The halo is not present in the 
 F172M image, nor in any of the NUV images or even in the optical images. This situation is
 similar to that in NGC 40, NGC 6302 and NGC 2440. The FUV emission is very likely,
  the result of H2 molecular fluorescent emission from the AGB ejecta, from molecules
 excited by the diffuse UV radiation from the CSPN seeping through to the cold molecular 
 region. Inspite of the presence of a very hot central star, with T$_{\rm eff}$ of 115000 K  
  (Toal{\`a} et al. 2014 ), and an earlier excursion to hot PN stage (born-again), the nebula 
  seems to still possess some unionized molecular gas. Is this gas a survivor of 12000 years 
  of the PN evolution? The FUV halo seem to be distributed asymetrically only on one side of 
  the nebula. Earlier, Dinnerstin \& Lester (1989) discovered  an infra-red disk inside the 
  nebula. A possible connection of this dust disk to the FUV halo needs to be explored.
  
\begin{figure*}
\vspace{0.3cm}
\includegraphics[width=8.0cm,height=4cm]{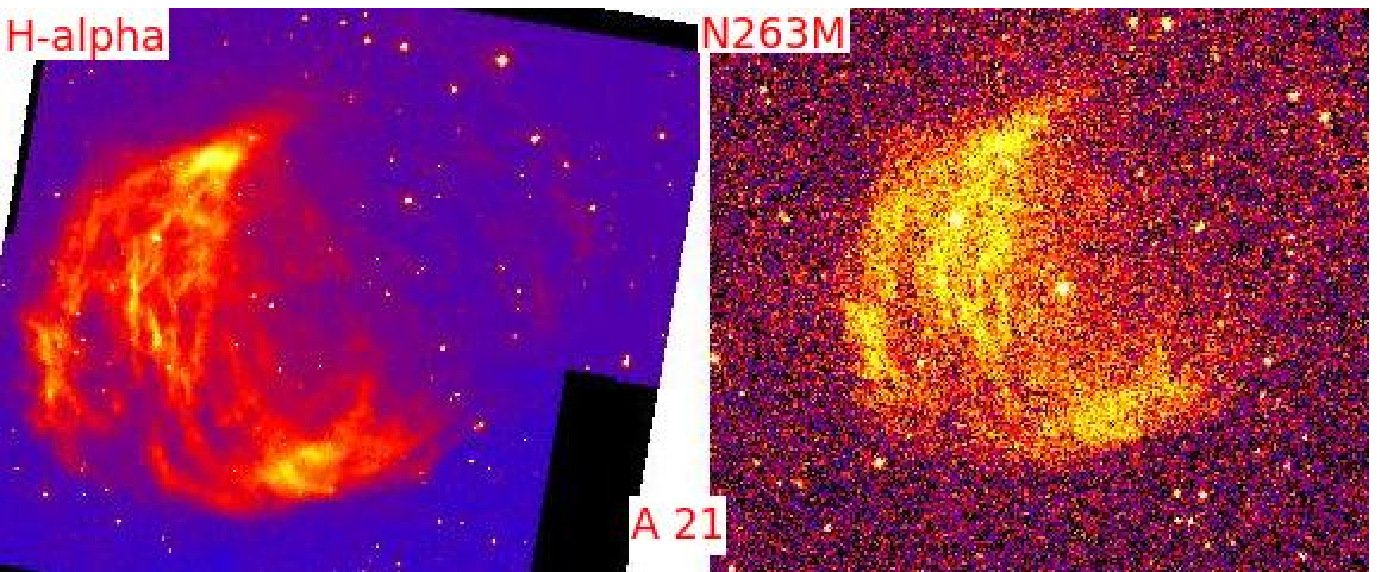}
\vspace{0.3cm}
\includegraphics[width=8.0cm,height=4cm]{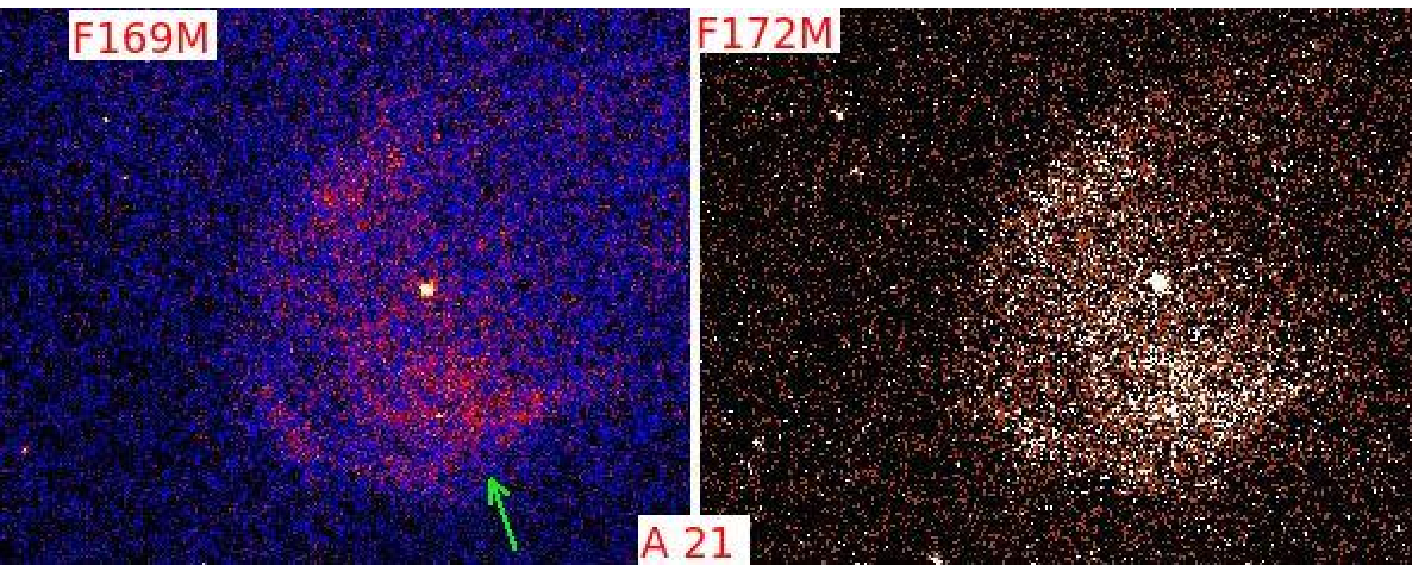}
\vspace{0.3cm}
\includegraphics[width=8.5cm,height=6.5cm]{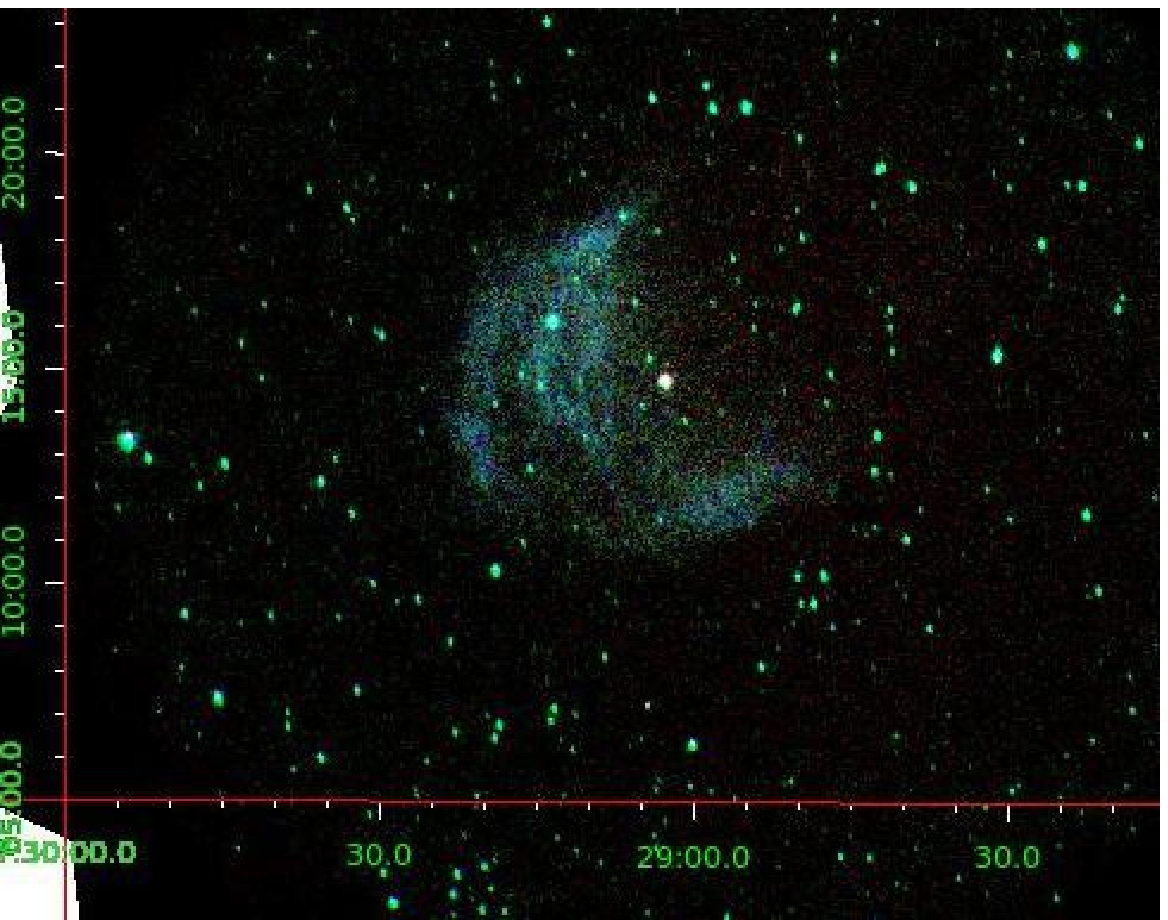}
\vspace{0.3cm}
\caption{Top left panel: Images of A 21 in H$\alpha$ (Manchado et al. 1996) and in N263M
 showing the similarity of the filamentary structure. right panel: The FUV images in
 F169M and F172M. The arrow points to the faint filament that connects to the CSPN.
 Bottom:  UVIT  composite image of A 21 in three colours F169M (blue)
 N263M (green) N245M(red). }
\end{figure*}

\subsubsection{ NGC 3587 -The Owl Nebula}

                This is a well studied PN with an angular size of $\sim$3'
  and of symmetrical morphology consisting of triple shell structure with a
  round double shell which forms the main bright nebula, and a faint outer halo (Chu et al. 1987; 
 Kwitter et al. 1992, Guerrero et al. 2003). The bright inner shell is about 
 $182" \times  168"$ and resembles the face of the owl with each eye being of $\sim 35"$. The
 outer shell is almost circular with a diameter of $\sim 208"$. The outer shell is about
 25\% larger than innershell. The surface brightness of the outer shell
 decreases outward in H$\alpha$ and [O\,{\sc iii}]. It shows a limb brightening
 along the PA -15$^\circ$ to 13$^\circ$ and in PA 180$^\circ$ to 230$^\circ$ 
 (Custeo et al. 2000).

               The halo is prominent in [O\,{\sc iii}] images than in H$\alpha$ or 
 [N\,{\sc ii}] but is present in all the optical  images (Hajian et al. 1997). This
  behavior is also common to other PN halos and is most likely an effect of the 
  hardening of ionizing radiation (Guerrero \& Manchado 1999). At the faintest level the
  halo is circular with a radius of 350" (Guerrero et al. 2003) although overall
  morphology is asymmetrical (Figure 13). The halo is kinematically independent of the main nebula
  (Chu et al. 1998). The relative line strength in the halo are also different from the
  main nebula. The [N\,{\sc ii}]/H$\alpha$ ratio of the halo is a factor of 4 higher
  than the other parts of the main nebula which indicates that the halo is photo-
  ionized. The morphology of the halo suggests
 an interaction with the surrounding interstellar medium and the gas in the halos is
ionized by stellar UV radiation leaking through the material of the main nebula(Guerrero
  \& Manchado 1999, Guerrero et al. 2003) 
                  
               IUE spectra obtained 1.5' away from the central star, almost
  into the halo, still show weak He\,{\sc ii} 1640\AA\ and C\,{\sc iii}] 1909\AA\. The halo
  gas is clearly ionized but of low density. The hot CSPN with $T_{\rm eff} = 104000$ K and
 $\log_g = 7.0$ shows weak stellar wind (Werner et al. 2019, Garc{\'i}a-D{\'i}az et al. 2018). A detailed 
  spatiokinematical study was presented by Garc{\'i}a-D{\'i}az et al. (2018), who treat the Owl
  kind of nebulae as a seperate class. One of the puzzles presented by the Owl is the
  existance of cavities (the eyes of the owl) and how are they maintained.
  At a Gaia distance of 880 pc the halo extends to 0.4 pc from the central star. Halo 
  expansion velocity is assesed as about $10$ km $^{-1}$ (Guerrero et al. 2003) which suggests
  an age $> 40,000$ years. The time scales indicate that the main nebula consists 
  of super-wind from AGB phase and the faint halo from an early AGB wind.

                Our observation with UVIT consists of FUV images in F154W, F169M, and F172M.
  We coupled these with the NUV image from GALEX (GI6-015012-PK148p57d1pp-nd-int), and compare them 
with image in the optical narrowband H-$\beta$ and N\,{\sc ii} filters.
  Images in all filters show the same general features as in the optical -- i.e.. the two shells, and halo.
   However, the cavities  (the eyes) are not as prominent in the FUV as in optical.
   Moreover, the NUV image shows a sharp wiggly boundary on the
  north-east showing the interaction with ISM in the direction of motion. The south
  western side is more diffuse and faint but more extended than NE. The FUV F169M image
  possibly dominated by He\,{\sc ii} emission, shows a more diffuse but brighter halo that extends all around
  the outer shell similar to NUV. The asymmetrical distribution of the halo does suggest
  interaction with ISM.

                One of the special features seen in the F169M image is a jet in the north-west
   cavity (see Figure 13) that is not seen either in F154W or F172M or even NUV images. This could be 
   a hot He\,{\sc ii} emitting jet, a real sign of activity in a docile nebula.

 \subsubsection{EGB6: }                   

                 EGB 6 (PN G221.6+46.4) is a large ($13' \times 11'$) and very low-surface-brightness 
  planetary nebula,  serendipitously discovered by Bond on POSS prints in 1978.
  The central star,  PG 0950+139  is a very hot DAOZ white dwarf with
  $T_{\rm eff} = 105000 \pm 5000$ K, $\log_g=7.4 \pm 0.4$ (Werner et al. 2018). The CSPN
   has an apparent cool dwarf companion shrouded in dust
   at a separation of 0."16, which was initially detected through near
   infrared excesses (Bond et al. 2016). Initial spectroscopic observations 
  showed the central star has strong [O\,{\sc iii}] emission. Later
 Liebert et al. (1989) showed that the strong [O\,{\sc iii}] nebular lines arise from a
compact emission knot (CEK), which is unresolved and appears to coincide with the PNN 
in ground-based images. However, recent HST observations (Bond et al. (2016))  showed that
 even the emission knot is associated with a companion at 0."166 away from the CSPN. 
 This corresponds to a projected linear separation of $\sim$118 AU, for a nominal distance
of about 725 pc. The electron density of the CEK is remarkably high, about $2.2 \times 10^{6}$ cm$^{-3}$, 
according to an emission-line analysis by Dopita \& Liebert (1989).

Thus, EGB 6 raises several astrophysical puzzles, including
how to explain the existence and survival of a compact dense
[O\,{\sc iii}] emission nebula apparently associated with a cool M
dwarf, located at least 118 AU from the source of ionizing
radiation (Bond et al. 2016). It is to be noted that very weak [O\,{\sc iii}] $\lambda$5007 emission 
attributed to the large PN has been detected serendipitously in the SDSS spectra of two faint galaxies that happen
to lie behind EGB 6 (Yuan \& Liu 2013). Acker et al. (1992) list the relative
intensities of H$\alpha$, H$\beta$, and 5007\AA.

 Bond et al. (2016)  suggest a scenario in
which the EGB 6 nucleus is descended from a wide binary similar to the Mira system,
 in which a portion of the
wind from an AGB star was captured into an accretion disk around a companion
 star; a remnant of this disk has survived to the present time and is surrounded
 by gas photoionized by UV radiation from the WD.

            Our UVIT observation include imaging EGB 6 in F148W, F154W, F169M and F172M
 filters. The low surface brightness PN is present in all the filters with about same
 dimensions and appearance as the H$\alpha$ and [N\,{\sc ii}] images (Jacoby et al. 1995).
  The  south-west part of the nebula is the brightest and even the F148 image shows  
   wavy appearence which coincides with  proper motion direction of the star.

              The most interesting and unique structure that is only present in FUV
  filters F148W, F154W and F169M is a smaller nebula close to the CSPN extending to about
  2.'4 away from it. It has a bow like appearence away from the central star (Figures
  14, 15) in the general  direction of proper motion of the star. This central nebula
  is not even present in Galex image. The existence of 
  this small nebula adds  a new puzzle to the already listed ones by Bond et al. (2016).

\subsection{Elliptical Nebulae }

             The members of this group from Table 1 are A 21, Jr Er 1, LoTr5, A70 ,Hu 1-2
  and NGC 7293. Reduced level 2 data of LoTr5,, Jr Er1, A 70, Hu 1-2 is not available. We
  present the results on thee PNs A21, and NGC 7293 in the following sections.

\subsubsection{ Abell 21, Medusa Nebula, A21:}   

               A21 (PNG205.1+14.2) is a very wide (685" x 530") evolved PN with a 
 very hot white dwarf central star WD0726+133 also known as YM29. The T$_{\rm eff}$ 
 and log g of the CSPN are estimated to be  140000 K and 6.5 respectively. 
  
 The WD central star appears as a point source superposed on diffuse emission in the MIPS
 24$\mu$m image (Chu et al. 2009). The flux density in the
MIPS 24$\mu$m band is almost three orders of magnitude higher
than the expected photospheric emission.  However, WD 0726+133 remains as a puzzle
since no companion has been detected (Ciardullo et al. 1999). The IR excess is attributed
 to a dust disk. The orgin of the dust is unclear  whether it is accreted material or 
 remnants of the dusty AGB phase. Whether the star is a binary is also a possibility (Clayton 
 et al. 2014). A21 shares this property with JrEr 1 and NGC 7293.
                           
                Short exposure UVIT observations have been obtained in F169M and F172M in
 FUV and N263M and N245 in NUV filters. The central star as expected is conspicuous. 
  The images are very similar to the optical ones. The NUV images show great similarity to 
   H$\alpha$ image with a large number of filaments (Figure 16). However the FUV images look more
  diffuse similar to [O\,{\sc iii}] $\lambda$5007 image (Manchado et al. 1996). 
  The F169M (includes He\,{\sc ii} 1640\AA\ and C\,{\sc iv} 1540\AA\ )  image shows a 
 faint filament connecting the central star. However better observations are required.

\begin{figure*}
\centering
\begin{minipage}{120mm}
\includegraphics[width=8cm,height=5cm]{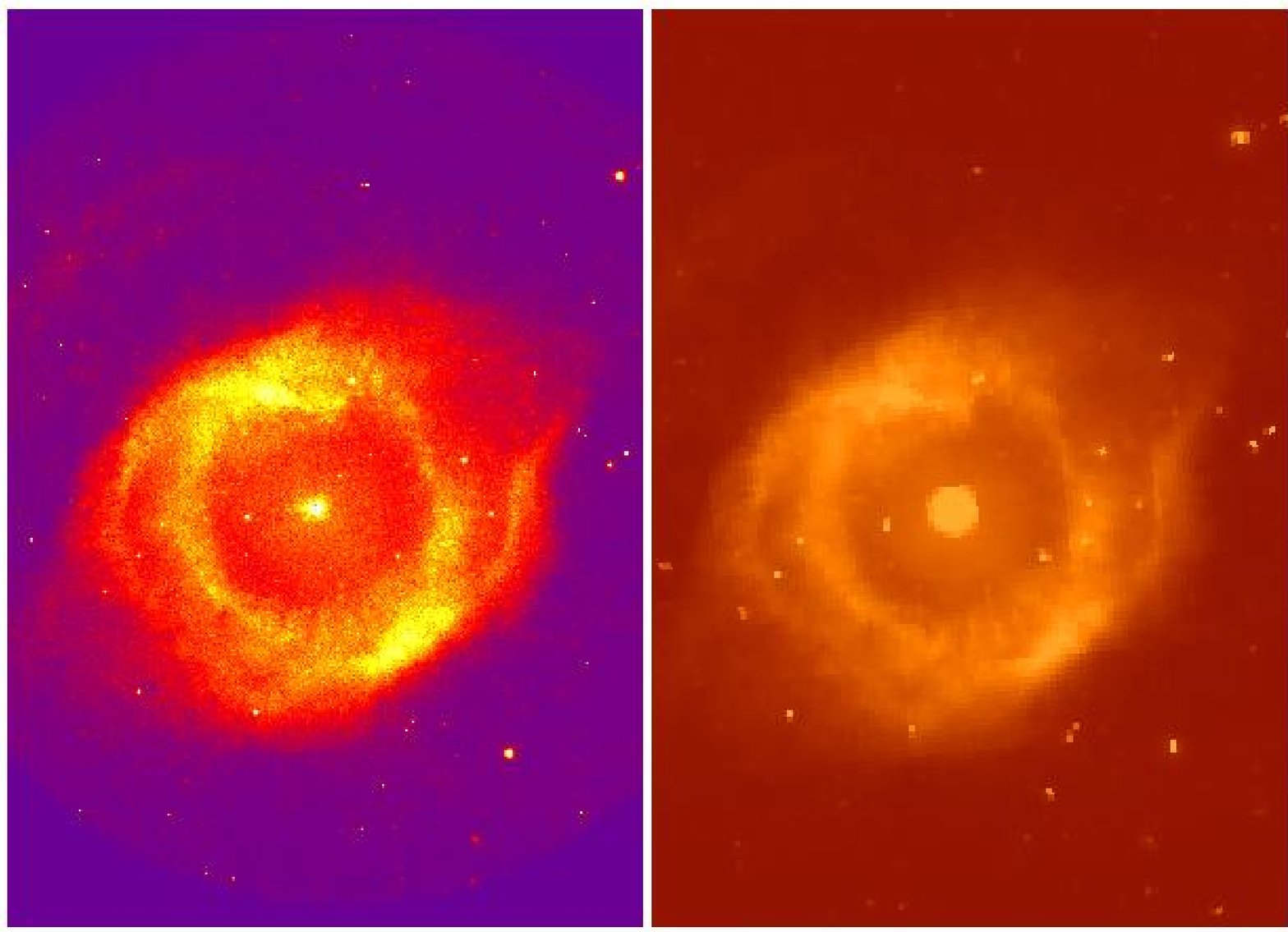}
\end{minipage}\\
\begin{minipage}{120mm}
\includegraphics[width=8cm,height=5cm]{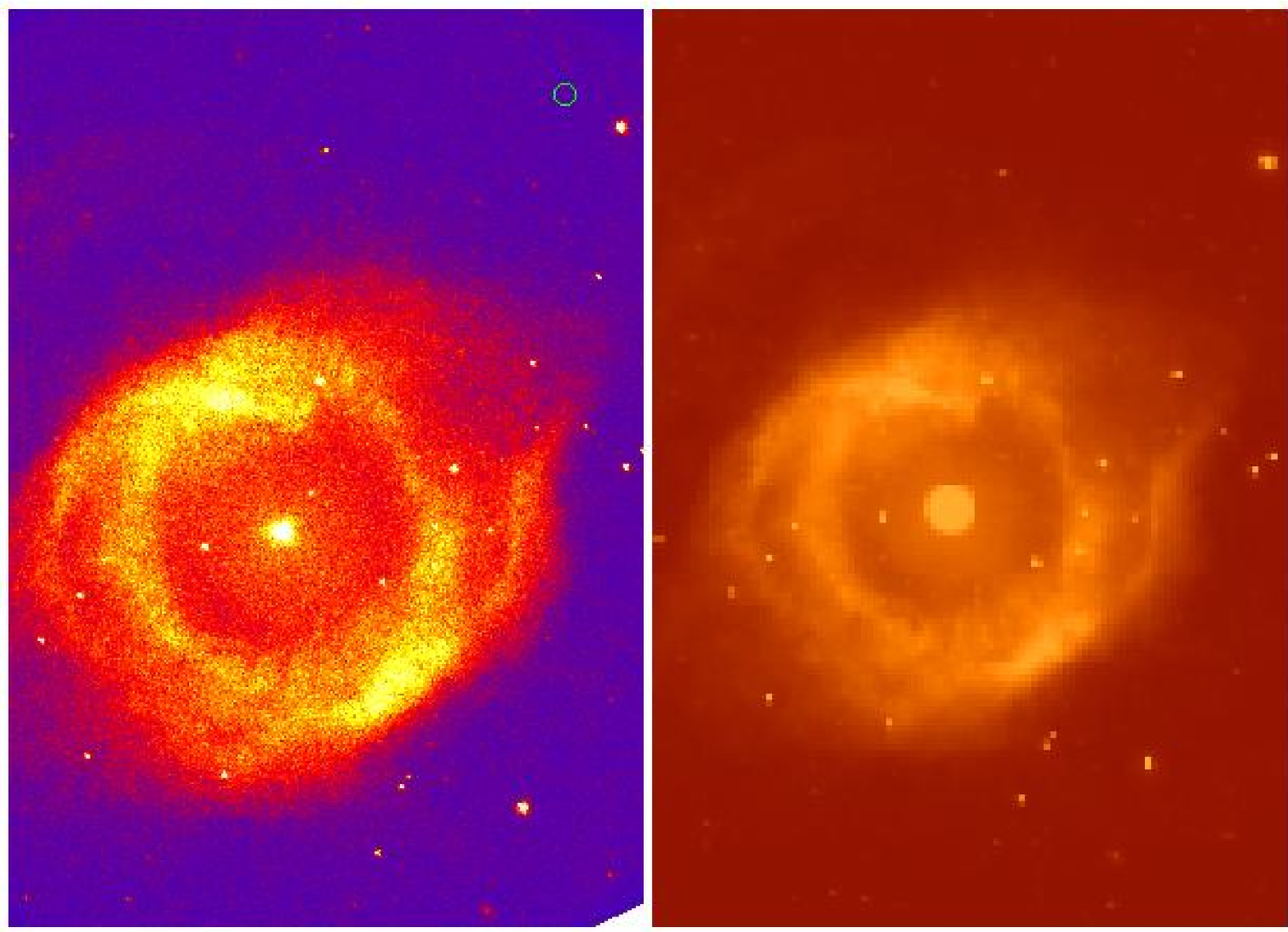}
\end{minipage}
\caption{ NUV image of NGC7293 in N245M (left) compared with Galex NUV image. Image
 in N263M (right)is again compared with NUV Galex image of NGC 7293. 
  Note the highier spatial resolution of UVIT images and  the radial
  filments.}
\end{figure*}


\begin{figure*}
\includegraphics[width=12cm,height=8cm]{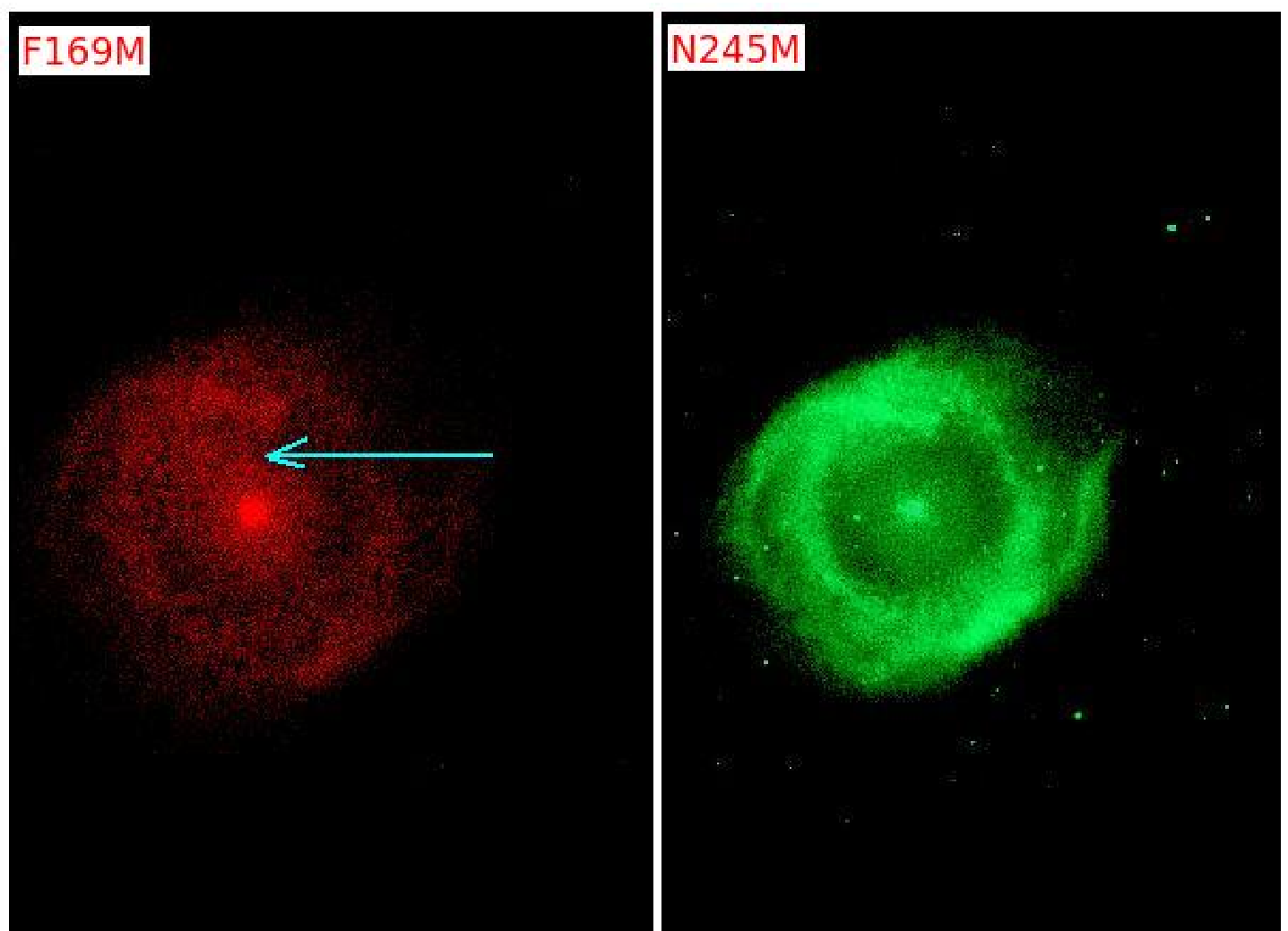}\\
\includegraphics[width=12cm,height=5cm]{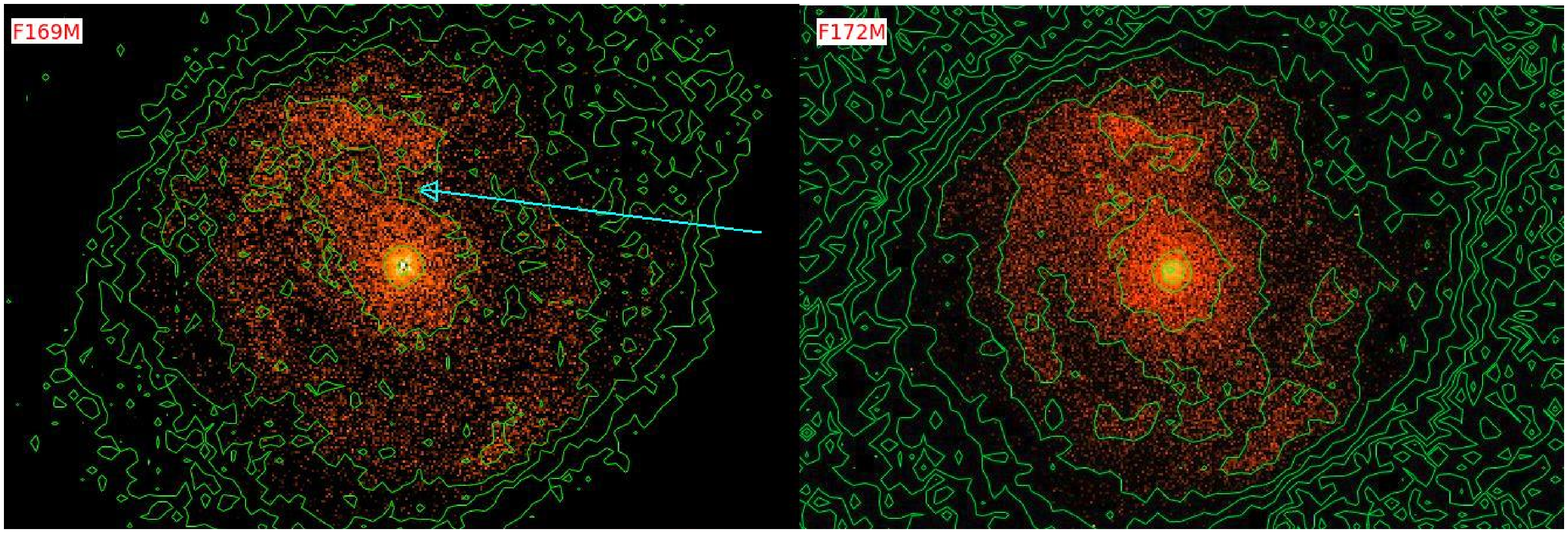}
\caption{ Top, 2 Panels: UVIT image of NGC 7293 in F169M filter (red) showing
the  outer nebula and the He\,{\sc ii} filament (marked by arrow) connecting the central
 region (and the star). The top right panel shows the absence of He\,{\sc ii} filament
 in the N245M image. 
  Bottom 2 Panels: Inner 4' regions of NGC 7293 in F169M (left) and F172M (right) superposed
  with intensity contours. The He\,{\sc ii} filament is seen in the F169M image (marked by
 arrow) where as it is absent in the F172M image.  }
\end{figure*}

\begin{figure*}
\vspace{0.0cm}
\includegraphics[width=9cm,height=6cm]{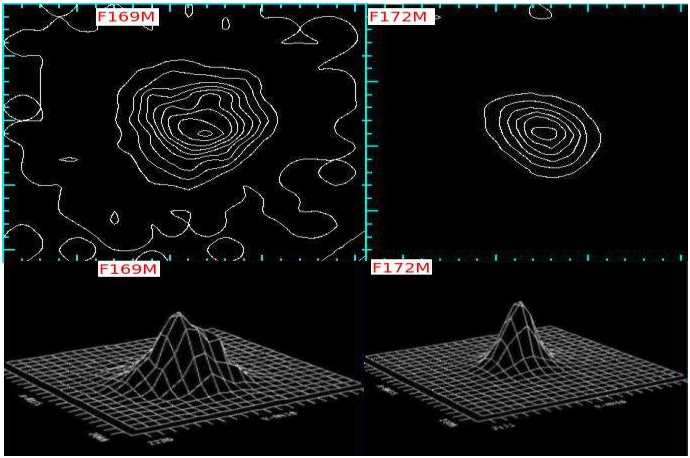}
\vspace{0.3cm}
\caption{A surface and contour study of NGC 7293 in UVIT/FUV  F169 and F172M bands.  
 Top: Surface contours of the central source in F169 (left) and F172M (right) showing
  distored contours for F169M indicating the possible presence of He\,{\sc ii} bright clouds or clumps
  around the CSPN. In F172M the contours more circular and suggests the presence of
  CSPN alone. 
  Bottom: The figure shows the volume contours of the central source in F169M (He\,{\sc ii}
 emission line) and F172M (continuum) filters. The CSPN is seen to be accompanied by three He\,{\sc ii} clumps in F169M 
  where as the profile in F172M shows CSPN alone. }
\end{figure*}


\begin{figure*}
\includegraphics[width=10cm,height=10cm]{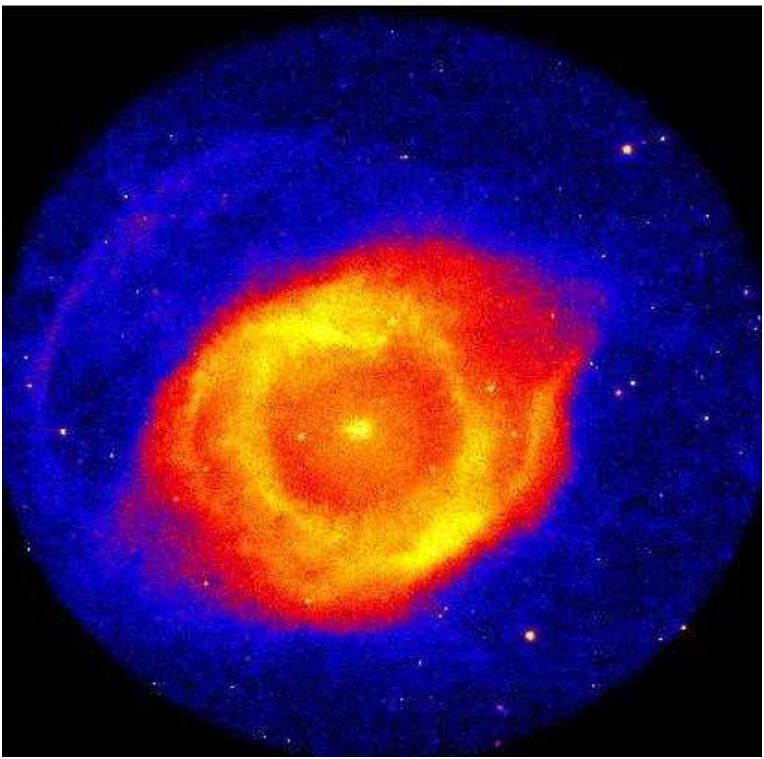}
\caption{ Details of UVIT image of NGC 7293 in the N263M filter. There are some new structures that
 are not seen in optical images.}
\end{figure*}


\begin{figure*}
\includegraphics[width=17cm,height=7cm]{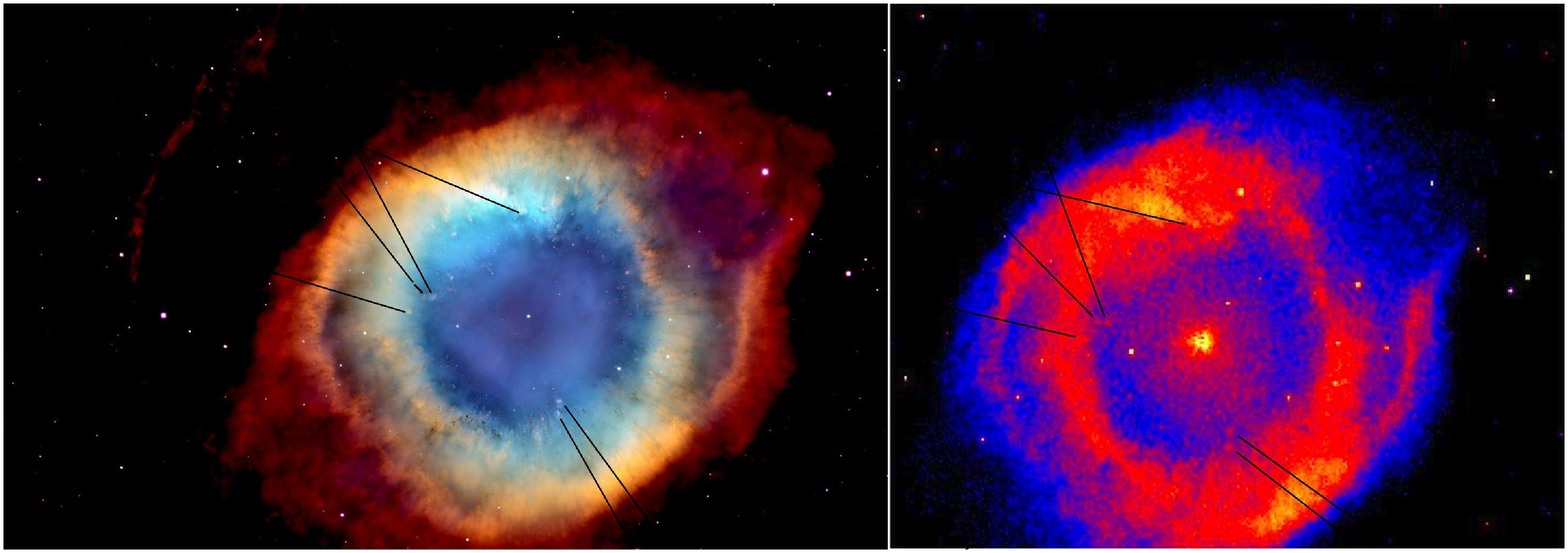}
\caption{ Details of UVIT image of NGC 7293 in the N263M filter. The fine optical knots detected
 in HST image (left) (O'Dell 2005) are seen UV image in F263M (right) obtained with UVIT. 
 The same knots
 are shown in  both images. Some of the cometary knots are also present in UV images.}
\end{figure*}

\subsubsection{NGC 7293, Helix Nebula:}

                           Helix Nebula (PNG036.1-57.1) is one of the nearest (Gaia distance
    of 201 pc) and well studied PN in almost all wavelengths (O'Dell et al. 2004,2005, 2007;
    Meaburn et al.  2005,2008; Hora et al. 2006; Mentez et al. 2015; Van de Steene et al. 2015). It is also
  the  largest PN in the sky with a diameter of about 13.'5. A Wide verity of phenomena has been 
    studied in this PN ranging from exotic molecules,  to X-rays, number of intriguing structures, 
    from small cometary knots to large-scale arcs, to bipolar outflows, dusty disks, shock fronts etc. 
     The inner helical structure is composed of thousands of cometary knots of lowly ionized and molecular gas 
   (O'Dell et al. 2007, Etxaluze et al. 2014).

      The white dwarf central star WD 2226-210 with a surface temperature of
        103600$\pm$ 5500 K (Napiwotzki 1999) ionizes the AGB nebula.
         Su et al. (2007) also showed the presence of a 35-150 AU diameter
       debris disk around this central star.

    The main nebula consists of two rings of highly ionized gas and a faint outer filament. 
   The three-dimensional (3D) structure of the main nebula has been investigated by  
    Zeigler et al. (2013)  who noted that the structure of the Helix
              projects as if it were a thick walled barrel composed of red and
             blue-shifted halves in a bipolar geometry. The barrel axis
         of the Helix is tilted about 10$^\circ$ east and 6$^\circ$ south relative to the
     line of sight .

                           There are two intriguing aspects that were revealed by infrared 
   studies.  Strong emission lines of [O\,{\sc iv}] 25.9$\mu$m and [Ne\,{\sc v}] 24.3
   $\mu$m have been detected in the Spitzer spectrum. This source of emission is identified
   with a point soure centred on the CSPN. The excitation of [O\,{\sc iv}] requires photons of
   54.9 e.v . In such a case, He\,{\sc ii} lines which also needs 54.4 e.v for ionization 
   (and recombination)are 
   expected to produce a strong  point source centered on the star. Images of He\,{\sc ii} 
    $\lambda$4686 have not shown such a point source. Secondly it has been known that
    Chandra X-ray imaging showed a  hard X-ray point source, even at subarc second resolution,
   centered on the CSPN. The source of X-rays is unknown. The surface temperature of CSPN 
    is not hot enough to generate hard X-rays from its photosphere. FUV region contains
    He\,{\sc ii} 1640\AA\ line, which is expected to be about 16 times stronger
    (recombination) than He\,{\sc ii} $\lambda$4686, occurs in the F169M filter of UVIT.
    With a view to map the hot gas at higher spatial resolution (1."3) we observed the
    central regions of Helix nebula.  

                        We observed the central region of Helix in the first instance in
     two FUV filters F169M and F172M and two NUV filters N245M and N263M.  F172M images
      expected to provide the continuum emission in contrast to F169M. NUV filters might
    include mostly the nebular continuum inaddition to some low excitation emissions.
    IUE spectra up to 2 arc minutes away from CSPN show that He\,{\sc ii} 1640\AA\ line provides 
     strongest emission in the inner regions. Figure 17 show the relative comparision 
     of the spatial resolution of UVIT with respect to Galex both in FUV and NUV.
     Several nebular knots (cometary) seen in HST optical images (O'Dell 2007) are also
     present in UV (Figure 20, 21).

                        F169 images of the central region of Helix showed mainly two surprising 
     features in contrast to F172M (and N245M and N263M). The region around CSPN has 
     nebulous clumps surrounding the point source (Figure 17). The intensity contours
     around the central source show few arc seconds extended regions suggesting that the
     central source is surrouned by clouds of He\,{\sc ii} (Figure 19). Secondly, there
     is a nebulous streamer connecting the central region to the outer ring (Figure 18). The
     He\,{\sc ii} streamers could possibly be providing mass flow to and from the central source. Could this be
     the source of accretion to the central WD?

      Although NGC 7293 is of great astrophysical interest, it is not
            an easy object to observe because of its low surface brightness
          and its large angular extent in the sky. The field of view of UVIT 
         is adequate to cover major portions of the nebula at any given  pointing.              
             We have observed two more locations in NGC 7293 covering the whole
     nebula including the shocked regions. Some of the data is not yet available. We hope
     to present  a detailed paper later.

\section{Concluding remarks:}

             In this paper, we have presented UVIT observations of 11 of the 19 proposed objects in our program. 
    This compendium forms an "atlas" of sorts, of deep UV imaging of thise objects, with a spatial resolution of 1".5.
    The data for the 8 objects is not yet available. The main theme that has been developed here is existence of  the  
    FUV haloes. Halos around PN are well  known and well studied (Ramos-Larios \& Phillips 2009). E.g., in NGC 3587, 
    the halo is seen in all wavelengths from UV (sec.2.2.3) to the  optical, and consists mainly of
     ionized gases (Guerrero \& Manchado  1999).  Extended ionized halo has been found around
    60\% of the PNs for which proper imaging has been done (Corradi et al. 2003).
   The halos are thought to be a result of mass-loss at the end of the AGB phase, their edges
    being the signature of the last thermal pulse (Shteffer \& Schonberner 2003).
   In contrast,  the UVIT discovered FUV halos,  and jets around bipolar nebulae and A 30 are
    only seen in wavelengths shortward of $\lambda$1650 not in longer wavelength images.    
    Warner and Lyman bands of H$_{2}$ start appearing shortward of $\lambda$1650.
    Ultraviolet fluoresence spectra of H$_{2}$ as modeled by France et al. (2005) with
    IC 63, show strong emission peak at $\lambda$1608 ($\lambda$$_{\rm eff}$ of F169M filter)
    and no emission shortward of $\lambda$1650.
    
               UVIT studies have brought out a totally  new aspect to the  hidden mass  of the
     planetary nebulae namely existence of  FUV halos, jets and arcs, mostly due to very cold 
    H$_{2}$ gas around young, bipolar and even some old PNs like born-again PNs as well. Such cold
     gas could only be seen through UV fluorescence of H$_{2}$ molecule. Big FUV lobes, and jets
     much bigger than optical nebule have been detected through FUV studies by UVIT.

     UV imaging in each case of PNs we studied, revealed a new aspect observationally  which 
    reiterates the importance of UV studies.

                      `Planetary nebulae are like a box of chocolates, you never know what
  you are going to get'  -- Monic Yourcg

\section{Acknowledgements:}  We would like to express our sincere thanks to several people
    who have helped us in carrying out this study of PNs, especially K. Sriram,
Arturo Manchado for supplying ground based images of some the PN in nebular lines, Don Goldman for the colour image of EGB 6, and Annapurni Subramaniam for the contour images of NGC 7293. 

This publication uses the data from the AstroSat mission of the Indian Space 
Research Organisation (ISRO), archived at the Indian Space Science Data Centre
(ISSDC).

  UVIT and ASTROSAT observatory development took about two decades before launch.
 Several people from several agencies were involved in this effort. We would like to thank them
 all collectively. NKR  would like thank Department of Science and
  Technology for their support through grant SERB/F/2143/2016-17 `Aspects in Stellar and Galactic Evolution'. 

  Some of the data presented in this paper (eg. Galex, IUE, HST) were obtained from the Mikulski
 Archive for Space Telescopes (MAST). STScI is operated by the Association of
 Universities for Research in Astronomy, Inc., under NASA contract NAS5-26555.
 Support for MAST for non-HST data is provided by the NASA Office of Space
 Science via grant NNX09AF08G and by other grants and contracts.

\begin{landscape}
\centering
\tabularfont
\begin{minipage}{200mm}
Target characteristics, and observation log of the PNs selected for our program. The morphological classification is as follows: Compact Bipolar nebulae (B), large Elliptical (E) and round (R). Subscipts "s", "m" and "a" refer to inner structure, multiple shells and ansae respectively. Objects that have been proposed but not observed to date are noted in the last column as "n.d.".
\begin{footnotesize}
\begin{tabular}{lcrcrrrrrr}
\hline
 Nebula   &\multicolumn{2}{c}{$\alpha$,$\delta$(2000)} &          &        &  \multicolumn{3}{c}{}&\multicolumn{2}{c}{ } \\
\cline{2-3}   \\
          &                   &                         & type (a) & size   & proposal  & Date &  FUV    & NUV      &  Comments  \\
           & h m s  &  $\circ$ ' "                     &          & arcmin & ID       &      &  filter & filter   &    \\
\hline
NGC   40 & 00 13 01.0 & +72 31 19.1 &  B  & 0.61$\times$ 0.61       & G06$_{-}$065 & 2016.12.09 & F3,F5 &B4,N2,Gr&  A\&A \\
NGC  650 & 01 42 19.7 & +51 34 31.5 &  Br & 3.2 $\times$ 1.97       & A07$_{-}$059To1&2019.10.04& F2,F3,F5&  &n.d.  \\
NGC 1514 & 04 09 17.0 & +30 46 33.5 &  R  & 1.67 $\times$ 1.67      & G06$_{-}$066&2016.12.26 &  F2,F3,F5&B4,B13,N2&     \\
   A 21  & 07 29 02.7 & +13 14 48.4 &  Es & 10.25 $\times$ 10.25    & G05$_{-}$178T01&2016.09.30&  F3,F5 &B4,B13 &       \\
         &           &           &     &           &A07$_{-}$134T02 & 2019.11.04& F2,F3,F5&      &       \\
NGC 2440 & 07 41 54.9 & -18 12 29.7 &  B,M & 0.55 $\times$ 0.55     & G07$_{-}$034 &2017.04.04& F3,F5 & B4,B13&      \\
OH231.8+4.2 & 07 42 16.9 & -14 42 50.2 &B  & 0.5 $\times$ 0.83      & A05$_{-}$103&2019.03.03& F3,F5 &  &n.d,ppn Calabash\\
JrEr1    & 07 57 51.6 & +53 25 16.9 &  Es & 6.3 $\times$ 6.3        & G06$_{-}$061 &2018.01.22& F2,F3,F5&  &PNG164.8+31.1\\
A 30     & 08 46 53.5 & +17 52 46.8 &  Rs & 0.29 $\times$ 0.13      & G06$_{-}$068 &2016.12.26& F2,F3,F5&B15,N2,Gr&  Hyd.def\\
NGC 2818 & 09 16 01.5 & -36 37 37.4 &  B  & 0.67 $\times$ 0.67      & A05$_{-}$149&2018.12.21& F2,F3,F5,Gr&  &    A$\&$A(p)\\
         &           &           &     &           &A09$_{-}$047    & 2020.06.09& F1,F2,F3,F5&  &   (d.n.r)   \\
EGB6     & 09 52 59.0 & +13 44 34.9 & R   & 0.16 $\times$ 0.16      & G08$_{-}$029 &2018.04.04& F1,F2,F3,F5&B4,B13,N2,B15&    \\ 
NGC 3587 & 11 14 52.8 & +55 02 00.0 & Rsm & 3.5 $\times$ 3.5        & A05$_{-}$149T03&2018.12.22& F2,F3,F5 &     &       \\
Lo Tr5   & 12 55 33.8 & +25 53 30.6 & E   & 0.18 $\times$ 0.18      & G05$_{-}$182&2016.06.04&F3,F5 &B13,B4&0'.5x0'.5,n.d\\
MyCn18   & 13 39 35.1 & -67 22 51.7 &     & 0.21 $\times$ 0.21      & G09$_{-}$061T02&2020.06.06& F2,F3,F5 &  & n.d,Hour glass \\
Mz 3     & 16 17 15.0 & -51 59 42.2 & B   &                         & A05$_{-}$103 & 2019.07.26& F2,F3,F5 &  & n.d  \\
         &           &           &     &          &                 & 2019.06.07&         &     &         \\
NGC 6302 & 17 13 44.4 & -37 06 11.0 & B   & 1.0 $\times$ 4.85       & G06$_{-}$071    &2017.03.18 & F2,F3,F5 &B15,Gr,N2&A$\&$A \\
         &            &             &     &                         & A05$_{-}$103T06 & 2019.07.24 & F2,F3,F5 &  &n.d  \\
A 70     & 20 31 33.2 & -07 05 18.0 & E   & 0.8 $\times$ 0.65       & A07$_{-}$134T03&2020.05.14& F2,F3,F5 &  &n.d \\
NGC 7027 & 21 07 01.8 & +42 14 10.0 & B   & 0.5 $\times$ 0.4        & A05$_{-}$149T01&2018.10.05& F2,F3,F5 &  &   \\
Hu 1-2   & 21 33 08.3 & +39 38 09.5 & Ea  & 0.67 $\times$ 0.3       & A07$_{-}$059T05&2019.10.24& F2,F3,F5 &  & n.d \\
NGC 7293 & 22 29 38.5 & -20 50 13.6 & E   &13.4  $\times$ 13.4      & G05$_{-}$187 &2016.09.30& F3,F5  &B4,B13& Helix   \\     
         &           &           &     &          & G07$_{-}$31   &2017.07.24& F3,F5   & &position 2,n.d \\
         &           &           &     &          & A07$_{-}$134T01 & 2019.10.25& F3,F5& & Position 3,n.d \\
\hline
\end{tabular}
\\
$^{a}$: The types are taken from IAC Morphological Catalog of Northern Galactic Planetary Nebulae
        (Manchado et al. 1996).   \\
$*1$: n.d- data is not available, level2 data not received from ISSDC  \\
$*2$:Galex FUV size is bigger -7'\\
\end{footnotesize}
\label{default}
\end{minipage}
\end{landscape}


\begin{theunbibliography}{}
\vspace{-1.5em}
\bibitem{latexcompanion} Ackers, A., Ochsenbein, F., Stenholm, B., et al., 1992, `Strasbourg-ESO Catalogue of
 Galactic Planetary Nebulae' 
\bibitem{latexcompanion} Balick, B., 1987, AJ, 94, 671
\bibitem{latexcompanion} Balick, B., Frank, A., 2002, ARAA, 40, 439
\bibitem{latexcompanion} Balick, B., Frank, A., Liu, B., Huart-Espinosa, H., 2017, ApJ, 843, 208
\bibitem{latexcompanion} Bianchi, L., 2012, `Planetary Nebulae' ed.     IAU Symp., 283, 45
\bibitem{} Bond, H.E., Ciardullo, R., Esplin, T.L., Hawley, S.A., Libert, J., Munari, U., ApJ, 826, 139
\bibitem{} Borkowski, K.J., Harrington, J.P., Tsvetanov, Z., 1995, ApJ, 449, L143
\bibitem{} Borkowski, K.J., Harrington, J.P., Tsvetanov, Z., Clegg, R.E.S., 1993, ApJ, 415, 47
\bibitem{} Clayton, G.C., De Marco, O., Nordhaus, J., et al., 2014, AJ, 147, 142
\bibitem{} Chu, Y.-H., Chang, T.H., Conway, G.M., 1997, ApJ, 482, 891
\bibitem{} Chu, Y.-H., Ho, C.-H., 1995, ApJ, 448, 127
\bibitem{} Chu, Y.-H., Grundl, R.A., Conway, G.M., 1987, AJ, 116, 1882 
\bibitem{} Chu, Y.-H., Su, K.Y.L., Bilikov, J., et al. 2011, AJ, 142, 75
\bibitem{} Chu, Y.-H., Gruendl, R.A., Guerrero, M.A., et al., 2009, AJ, 138, 691
\bibitem{} Corradi, R. L. M., Schonberner, D., Steffen, M., Perinotto, M., 2003,  MNRAS, 340, 417
\bibitem{} Cox, P., Huggins, P.J., Maillard, J.-P., et al. 2002, A\& A,384,603
\bibitem{} Cuesta, L., Phillips, J.P., 2000, ApJ, 543, 754
\bibitem{} Cuesta, L., Phillips, J.P., 2000, AJ, 120, 266
\bibitem{} Delfosse, X., Kahane, C., Forveille, T., 1997, A\& A, 320, 249 
\bibitem{} De Marco, O., 2009, PASP, 121, 316
\bibitem{} Dinerstein, H.L.,  Lester, D.F., 1984, ApJ, 281, 702
\bibitem{} Dopita, M.A.,  Liebert, J., 1989, ApJ, 347, 910 
\bibitem{} Etxaluze, M., Cernicharo, J., Goicoechea, J.R., et al. 2014, A\& A, 566, 78
\bibitem{} Fang, X., Guerrero, M.A., Marquez-Lugo, R.A., Toala, J.A., et al., 2014, ApJ, 797, 100
\bibitem{} Feibelman, W.A., 2001, ApJ, 550, 785
\bibitem{} France, K., Andersson, B.-G., McCandliss, S.R., Feldman, P.D., 2005, ApJ, 628, 750
\bibitem{} García-Segura, G., Villaver, E., Langer, N., Yoon, S.-C., Manchado, A., 2014, ApJ, 783, 74
\bibitem{} Garcia-Diaz, M.T., Steffen, W., Henny, W.J., et al., 2018, MNRAS, 479, 3909
\bibitem{} Gusten, R., Wiesemeyer, H., Neufeld, D., 2019, Nature, 568, 357
\bibitem{} Guerrero, M. A.; Manchado, A., 1999, ApJ, 522, 378 
\bibitem{} Guerrero, M.A., Ruiz, N., Hamann, W.-R., et al. 2012, ApJ, 755,129
\bibitem{} Guerrero, M.A., Chu, Y.-H., Manchado, A.,  Kwitter, K.B., 2003, AJ, 125, 321 
\bibitem{} Hazard, C., Terlevich, R., Morton, D.C., Sargent, W.L.W., Ferland, G, 1980, Nature, 285, 463
\bibitem{} Hajian, A.R., Frank, A., Balick, B., Terzian, Y., 1997, ApJ, 477, 226
\bibitem{} Herald, J.E., Bianchi, L., 2011, MNRAS, 417, 2440
\bibitem{} Herwig, F., 2005, ARA\&A, 43, 435
\bibitem{} Hora, J.L. Latter, W.B., Smith, H.A., 2007, 652, 426
\bibitem{} Hsia, C -H, Chau, W., Zhang, Y., Kwok, S., 2014, 787, 25
\bibitem{} Hua, C. T., Dopita, M. A., Martinis, J., 1998, A\&AS, 133,.361
\bibitem{} Iben, I.Jr., Kaler, J.B., Truran, J.W., Renzini, A., 1983, ApJ, 264, 605
\bibitem{} Jacoby, G.H., 1979, PASP, 91, 754
\bibitem{} Jacoby, G.H.,  Van De Steene, G., 1995, AJ, 110, 1285
\bibitem{} Jones, D., Van Winckel, H., Aller, A., Exter, K.,  De Marco, O., 2017, A\&A, 600, 9
\bibitem{} Lang, D., Hogg, D.W., Mierle, K.,Blanton, M., Roweis, S., 2010, AJ, 139, 1762
\bibitem{} Latter, W.B., Kelly, D.M., Hora, J.L., Deutsch, L.K., 1995, ApJS, 100, 159
\bibitem{} Lawlor, T. M., MacDonald, J., 2006, MNRAS, 371, 263 
\bibitem{} Lester, D.F.,  Dinerstein, H.L., 1984, ApJ, 281, L67
\bibitem{} Liebert, J., Green, R., Bond, H.E., Holberg, J.B., et al., 1989, ApJ, 346, 251 
\bibitem{} Lopez, J.A., Vazquez,R., Rodriguez, L.F., 1995, ApJ, 455, L63
\bibitem{} Lopez, J.A., Meaburn, J., Bryce, M., Holloway, A.J., 1998, ApJ, 493, 803
\bibitem{} Kameswara Rao, N.,Sutaria, F., Murthy, J., Krishna, S., Mohan, R., Ray, A., 2018, A\&A, 609, L1
\bibitem{} Kameswara Rao, N., De Marco, O., Krishna, S., Murthy, J., Ray, A., Sutaria, F., Mohan, R., 2018, A\&A, 620, 138
\bibitem{} Kastner, J.H., Vrtilek, S.D., Soker, N., 2001, Apj, 550, 189 
\bibitem{} Kemper, F., Molster,F.J., Jager, C., Waters, L.B.F.M., 2002, A\&A, 394, 679
\bibitem{} Kumar, A., Ghosh, S.K., Hutchings, J., et al. 2012, in Proc. SPIE.Vol. 8443
\bibitem{} Kwitter, K.B., Chu, Y.H.,  Downes, R.A., 1993, IAUS, 155, 209 
\bibitem{} Kwok, S., Purton, C.R., Fitzgerald, P.M., 1978, ApJ, 219, L125
\bibitem{} Lago, P. J. A., Costa, R. D. D., 2016, RmxAA, 52, 329 
\bibitem{} Latter, W.B., Dayal, A., Bieging, J.H., Meakin, C., et al. 2000, ApJ, 539, 783
\bibitem{} Manchado, A., Stanghellini, L., Guerrero, M.A., 1996, ApJ, 466, L95
\bibitem{} Manchado, A., Guerrero, M.A., Stanghellini, L., Serra-Ricart, M., 1996, `The IAC Morphological
 Catalog of Northern Galactic Planetary Nebulae' ,IAC, La Laguna
\bibitem{} Mata, H., Ramos-Larios, G., Guerrero M.A., et al., 2016, MNRAS, 459, 841   
\bibitem{} Matsuura, M., Zijlstra, A.A., Molster, F.J., Waters, L.B.F.M., Nomura,
H., Sahai, R., Hoare, M.G., 2005, MNRAS, 359, 383
\bibitem{} Martin, C., Hurwitz, M., Bowyer, S., 1990, 354, 220
\bibitem{} Masson, C.R., 1989, ApJ, 336, 294
\bibitem{} Meaburn, J., Lopez, J.A., Steffen, W., Graham., M.F., Holloway,  AJ, 130, 2303
\bibitem{} Meaburn, J., Lopez, J.A., Bryce, M.,  Redman, M.P., 1998, A\&A, 334, 670 
\bibitem{} Meaburn, J., Lloyd, M., Vaytet, N.M.H., Lopez, J.A., 2008, MNRAS, 385, 269
\bibitem{} Miller, T.R., Henry, R.B.C., Balick, B., et al. 2019, MNRAS, 482, 278
\bibitem{} Miller Bertolami, M. M., Althaus, L. G., 2006, RMxAC, 26, 48 
\bibitem{} Montez, R. Jr., Kastner, .H., Balick, B., et al., 2015, ApJ, 800, 8
\bibitem{} Montez, R.,  Kastner, J.H., 2018, ApJ, 861,45
\bibitem{} Murthy, J., Rahna, P., Sutaria, F., et al., 2017, A\&C, 20, 120
\bibitem{} Napiwotzki, R., 1999, A\&A, 350, 101
\bibitem{} O'Dell, C.R., McCullough, P.R., Meixner, M., 2004, AJ, 128, 2339
\bibitem{} O'Dell, C.R., Henney, W.J.,  Ferland, G.J., 2005, AJ, 130, 1720
\bibitem{} O'Dell, C.R., Henney, W.J.,  Ferland, G.J., 2007, AJ, 133, 2343 
\bibitem{} Peretto, N., Fuller, G.A., Zijlstra, A.A., Patel, N.A., 2007, A\&A, 437,207
\bibitem{} Rahana, P.T., Murthy, J., Safonova,M., et al. 2017, MNRAS, 471, 3028
\bibitem{} Ramos-Larios, G.,  Phillips, J.P., 2009, MNRAS, 400, 575 
\bibitem{} Ressler, M.E., Cohen, M., Wachter, S., et al., 2010, AJ, 140, 1882
\bibitem{} Seaton, M.J., 1980, QJRAS, 21, 229
\bibitem{} Santander-García, M., Bujarrabal, V., Alcolea, J., Castro-Carrizo, A.,
 Sánchez Contreras, C., Quintana-Lacaci, G., Corradi, R. L. M., Neri, R., 2017,
 A\&A, 597, 27
\bibitem{} Schmidt, D.R., Ziury, L.M., 2016, ApJ, 817, 175
\bibitem{} Steffen, M.; Schönberner, D., 2003, IAUS, 209, 439 
\bibitem{} Su, K.Y.L., Chu, Y.-H., Rieke, G.H., Huggins, P.J., et al., 2007, ApJ, 657L, 41 
\bibitem{} Szyszka, C., Walsh, J.R., Zijlstra, A.A., Tsamis, Y.G., 2009, ApJ, 707,
 L32
\bibitem{} Szyszka, C., Zijlstra, A.A., Walsh, J.R., 2011, MNRAS, 416, 715
\bibitem{} Tandon, S.N., Hutchings, J.B., Gosh, S.K., et al., 2017a, JApA, 38, 28
\bibitem{} Tandon, S.N., Subramaniam, A., Girish, V., et al., 2017b, AJ, 154, 128 
\bibitem{} Tandon, S.N., et al., 2020, AJ, 159, 158
\bibitem{} Tarafdar, S.P.,  Apparao, K.M.V., 1988, ApJ, 327, 342
\bibitem{} Toal\'{a}, J.A., Guerrero, M.A.,  Todt, H., Hamann, W.-R., et al., 2015, ApJ, 799, 67
\bibitem{} Van De Steene G.C., Van Hoof, P.A.M., Exter, K.M., Barlow, M., et al., 2015, A\&A, 574, 134
\bibitem{} Vazquez, R., 2012, ApJ, 751, 116
\bibitem{} Walton, N. A., Pottasch, S. R., Reay, N.K., 1988, A\&A, 200, 21
\bibitem{} Wang,M.-Y., Muthumariappan,C., Kwok, S.,2006,  IAUS, 234,537
\bibitem{} Wang M.-Y., Hasegawa T.I., Kwok S., 2008, ApJ, 673, 264 
\bibitem{} Werner, K., Rauch, T., Kruk, J.W., 2018, A\&A, 616, 73
\bibitem{} Werner, K., Rauch, T., Reindl, N., 2019, MNRAS, 483, 5291
\bibitem{} Wesson, R., Cernicharo, J., Barlow, M.J. et al. 2010, A\&A, 518, 144
\bibitem{} Witt, A.N., Stecher, T.P., Boroson, T.A., Bohlin, R.C., 1989, ApJ., 336, L21
\bibitem{} Wright, N.J., Barlow, M.J., Ercolano, B., Rauch, T., 2011, MNRAS, 418, 370
\bibitem{} Zeigler, N.R., Zack, L.N., Woolf, N.J., Ziurys, L.M., 2013, ApJ, 778, 16 
\bibitem{} Zijlstra, A.A., Van Hoof, P.A.M., Perley, R.A., 2008, ApJ, 681, 1296
\end{theunbibliography}
\end{document}